\documentclass[11pt]{article}

\usepackage{microtype}%if unwanted, comment out or use option "draft"
\usepackage{xspace}
\usepackage{bbm}
\usepackage{times}
\usepackage{graphicx}
\usepackage{amsmath,amssymb,amsthm}

\usepackage{paralist}
\usepackage{bm}
\usepackage{xspace}
\usepackage{url}
\usepackage{prettyref}
\usepackage{boxedminipage}
\usepackage{wrapfig}
\usepackage{ifthen}
\usepackage{color}
\usepackage{xcolor}
\usepackage{algpseudocode}

\usepackage{algorithm}
\usepackage{fullpage}
\usepackage{hyperref}
\usepackage[compact]{titlesec}
\usepackage{cleveref}
\usepackage{nicefrac}

\usepackage[bottom=1in,top=1in,left=1in,right=1in]{geometry}

\usepackage{framed}

\usepackage{thmtools}
\usepackage{thm-restate}
\declaretheorem[numberwithin=section]{theorem}
\declaretheorem[sibling=theorem]{lemma}

\declaretheorem[sibling=theorem]{corollary}
\newtheorem{problem}[theorem]{Problem}
\newtheorem{definition}[theorem]{Definition}
\crefname{claim}{claim}{claims}

\usepackage{comment}
\newtheorem{proposition}[theorem]{Proposition}

\newtheorem{observation}[theorem]{Observation}

\usepackage[colorinlistoftodos,textsize=tiny,textwidth=2cm,color=red!25!white,obeyFinal]{todonotes}

\newcommand{\ignore}[1]{}

\newcommand{\sse}{\ensuremath{\subseteq}\xspace}

\newcommand{\initOneLiners}{%
    \setlength{\itemsep}{0pt}
    \setlength{\parsep }{0pt}
    \setlength{\topsep }{0pt}
}
\newenvironment{OneLiners}[1][\ensuremath{\bullet}]
    {\begin{list}
        {#1}
        {\initOneLiners}}
    {\end{list}}

\makeatletter
\newcommand{\setword}[2]{%
  \phantomsection
  #1\def\@currentlabel{\unexpanded{#1}}\label{#2}%
}
\makeatother

\usepackage{caption}

\DeclareCaptionFormat{algor}{%
  \hrulefill\par\offinterlineskip\vskip1pt%
    \textbf{#1#2}#3\offinterlineskip\hrulefill}
\DeclareCaptionStyle{algori}{singlelinecheck=off,format=algor,labelsep=space}
\makeatletter
\def\therule{\makebox[\algorithmicindent][l]{\hspace*{.5em}\vrule height .75\baselineskip depth .25\baselineskip}}%

\newtoks\therules% Contains rules
\therules={}% Start with empty token list
\def\appendto#1#2{\expandafter#1\expandafter{\the#1#2}}% Append to token list
\def\gobblefirst#1{% Remove (first) from token list
  #1\expandafter\expandafter\expandafter{\expandafter\@gobble\the#1}}%
\def\LState{\State\unskip\the\therules}% New line-state
\def\pushindent{\appendto\therules\therule}%
\def\popindent{\gobblefirst\therules}%
\def\printindent{\unskip\the\therules}%
\def\printandpush{\printindent\pushindent}%
\def\popandprint{\popindent\printindent}%

\algdef{SE}[WHILE]{While}{EndWhile}[1]
  {\printandpush\algorithmicwhile\ #1\ \algorithmicdo}
  {\popandprint\algorithmicend\ \algorithmicwhile}%
\algdef{SE}[FOR]{For}{EndFor}[1]
  {\printandpush\algorithmicfor\ #1\ \algorithmicdo}
  {\popandprint\algorithmicend\ \algorithmicfor}%
\algdef{S}[FOR]{ForAll}[1]
  {\printindent\algorithmicforall\ #1\ \algorithmicdo}%
\algdef{SE}[LOOP]{Loop}{EndLoop}
  {\printandpush\algorithmicloop}
  {\popandprint\algorithmicend\ \algorithmicloop}%
\algdef{SE}[REPEAT]{Repeat}{Until}
  {\printandpush\algorithmicrepeat}[1]
  {\popandprint\algorithmicuntil\ #1}%
\algdef{SE}[IF]{If}{EndIf}[1]
  {\printandpush\algorithmicif\ #1\ \algorithmicthen}
  {\popandprint\algorithmicend\ \algorithmicif}%
\algdef{C}[IF]{IF}{ElsIf}[1]
  {\popandprint\pushindent\algorithmicelse\ \algorithmicif\ #1\ \algorithmicthen}%
\algdef{Ce}[ELSE]{IF}{Else}{EndIf}
  {\popandprint\pushindent\algorithmicelse}%
\algdef{SE}[PROCEDURE]{Procedure}{EndProcedure}[2]
   {\printandpush\algorithmicprocedure\ \textproc{#1}\ifthenelse{\equal{#2}{}}{}{(#2)}}%
   {\popandprint\algorithmicend\ \algorithmicprocedure}%
\algdef{SE}[FUNCTION]{Function}{EndFunction}[2]
   {\printandpush\algorithmicfunction\ \textproc{#1}\ifthenelse{\equal{#2}{}}{}{(#2)}}%
   {\popandprint\algorithmicend\ \algorithmicfunction}%
\makeatother

\usepackage{subcaption}
\usepackage{tikz}
\usetikzlibrary{arrows.meta}
\usetikzlibrary{decorations.markings}

\usepackage{my_definitions_new}

\begin{document}
	\title{\permutation Strikes Back: The Power of Recourse in Online Metric Matching
	}
	\author{Varun Gupta\thanks{University of Chicago. Email: {\tt guptav@uchicago.edu}. Author is grateful for financial support from the University of Chicago, Booth School of Business.}
		\and Ravishankar Krishnaswamy\thanks{Microsoft Research India. Email: {\tt rakri@microsoft.com}}
		\and Sai Sandeep\thanks{Carnegie Mellon University. Email: {\tt spallerl@andrew.cmu.edu}. Work partly done at Microsoft Research India and partly supported by NSF grant CCF-1422045}
	}
	
	\maketitle
	\vspace{-0.3in}
	
	\begin{abstract}
In the classical \omm problem, we are given a metric space with $k$ servers. A collection of clients arrive in an online fashion, and upon arrival, a client should irrevocably be matched to an as-yet-unmatched server. The goal is to find an online matching which minimizes the total cost, i.e., the sum of distances between each client and the server it is matched to. We know deterministic algorithms~\cite{KP93,khuller1994line} that achieve a competitive ratio of $2k-1$, and this bound is tight for deterministic algorithms. Randomization can be used to overcome this lower bound, and we know $O(\log^2 k)$ competitive algorithms~\cite{bansal2007log} and an $\Omega(\log k)$ lower bound. The problem has also long been considered in specialized metrics such as the line metric or metrics of bounded doubling dimension, with the current best result on a line metric being a deterministic $O(\log k)$ competitive algorithm~\cite{raghvendra2018optimal}. Obtaining (or refuting) $O(\log k)$-competitive algorithms in general metrics and constant-competitive algorithms on the line metric have been long-standing open questions in this area.
			
In this paper, we investigate the robustness of these lower bounds by considering the Online Metric Matching with Recourse problem where we are allowed to change a small number of previous assignments upon arrival of a new client. Indeed, we show that a small logarithmic amount of recourse can significantly improve the quality of matchings we can maintain. For general metrics, we show a simple \emph{deterministic} $O(\log k)$-competitive algorithm with $O(\log k)$-amortized recourse, an exponential improvement over the $2k-1$ lower bound when no recourse is allowed. We next consider the line metric, and present a deterministic algorithm which is $3$-competitive and has $O(\log k)$-recourse, again a substantial improvement over the best known $O(\log k)$-competitive algorithm when no recourse is allowed. We finally illustrate another benefit of allowing limited recourse: we can extend the Online Metric Matching model to handle arrivals and departures of both clients and servers (as opposed to just handling arrivals of clients) and still maintain competitive solutions. Indeed, we show a simple randomized $O(\log n)$-competitive algorithm with $O(\log \Delta)$-recourse in this fully online setting, where $n$ is the number of points in the metric space and $\Delta$ is the aspect ratio of the underlying metric. Perhaps the most important technical contribution of this work is in showing that these improved results can, in fact, be achieved by suitably equipping a two-decades-old algorithm \permutation for online metric matching with limited recourse.
	\end{abstract}
% \tableofcontents

\setlength{\parskip}{.05in}
\thispagestyle{empty}

\maketitle

\clearpage
\setcounter{page}{1}

% !TeX spellcheck = en_US
\section{Introduction} \label{sec:intro}
The classical \omm problem is defined on a metric space $(\Xcal,d)$, where $\Xcal$ denotes a set of $n$ points where the servers or clients may be located, and a distance function (metric) $d : \Xcal \times \Xcal \to \RR_+$. A set $S \sse \Xcal$ of servers, $|S|=k$, is given offline, and then a sequence of client requests $C = (c_1, \ldots, c_k)$ is revealed in an online manner. The algorithm is required to match each client request to an available (previously unmatched) server on arrival, and this decision is irrevocable. The objective is to minimize the total cost of the matching, which is the sum of distances between each client and the server it is matched to. The quality of an algorithm is measured using competitive ratio, which measures the worst-case over all instances of the ratio of the cost of the online algorithm and the cost of an optimal offline matching of the instance. 

This problem was first considered in two independent works~\cite{KP93,khuller1994line} soon after the publication of the celebrated work of~\cite{kvv} on the online maximum matching problem. Both these works present a $(2k-1)$-competitive deterministic algorithm called \permutation, and also show that this bound is tight among deterministic algorithms.  The work \cite{meyerson2006randomized} show that randomization can overcome this lower bound (for oblivious adversaries) by giving a $\Ocal(\log^3 k)$-competitive randomized algorithm, which was subsequently improved to $\Ocal(\log^2 k)$~\cite{bansal2007log}. In contrast, the best known lower bound for randomized algorithms is a factor of $\Omega(\log k)$\cite{meyerson2006randomized}.  

The \omm problem has also elicited much interest in specialized metrics, such as the line metric and metrics of bounded doubling dimension. 
For \omm on a line (\omml), \cite{fuchs2005online} show a lower bound of $9.001$, disproving a long conjectured bound of $9$. Good algorithms had been elusive until recently, and the current best-known algorithm for \omml is a deterministic $\Ocal(\log k)$-competitive algorithm~\cite{raghvendra2018optimal}, prior to which the best-known results were an $\Ocal(\log k)$ upper bound for randomized algorithms~\cite{gupta2012online} and an $\Ocal(\log^2k)$ upper bound for deterministic algorithms~\cite{raghvendra2017}. There also exists an $\Omega(\log k)$ lower bound on natural families of algorithms for \omml~\cite{antoniadis2018collection,koutsoupias} making this an intriguing open problem.

Given these barriers for designing improved algorithms for \omm, we ask: \emph{can we obtain strictly better performance if we are allowed to re-match a few previous clients upon arrival of a new client?} 

\begin{problem} [\ommr]
	An instance consists of a metric space $(\Xcal,d)$, and a multi-set $S \subseteq X$ of servers with $|S|=k$. A sequence of client requests $C = (c_1, \ldots, c_k)$ is revealed in an online manner. At time $t$, after the algorithm observes $c_t$, it must maintain a matching $\Mcal_t$ such that every client is matched to exactly one server, and each server is matched to at most one client. The algorithm can re-match some earlier clients, and the number of clients re-matched is called the {\it recourse}.
\end{problem}

\begin{definition}
	We say that an online algorithm is $\alpha$-competitive with $\beta$-amortized recourse for \ommr if for all $t \in [k]$, the cost of the algorithm's matching for $C_t := (c_1, \ldots, c_t)$ is at most $\alpha$ times the cost of the optimal matching for $C_t$, and the total number of recourse steps taken so far is at most $\beta t$. Additionally, the algorithm is said to have $\beta$-per-client recourse if no client is rematched more than $\beta$ times.
\end{definition}
While our main motivation is the theoretical understanding of the power of limited recourse in the classical \omm problem, often in practice it is also the case that matching/allocation decisions are not irrevocable and there is a cost (or) penalty for re-assignments. For example, in a live video streaming setting, the users arrive online and want to stream a video, and the ISP must choose a server to stream from preferring a server closer to the user. Of course, this decision can be changed over the time horizon, but this will cause a temporary disruption that must be minimized. %The other objective is to minimize the network distance of the client from the server it is matched to.
The recourse model then naturally captures the competing goals of minimizing cost as well as the number of re-assignments. The stronger notion of \textit{per-client} recourse additionally guarantees a fairness property by bounding the inconvenience to each client.   
%Another example (on a line metric) if of matching skis of different lengths to people of differing heights arriving at a ski rental facility. It is possible to re-match skis of people if it yields a better overall assignment, but is an avoidable action for the hassle it causes. 

In this paper, we present algorithms that highlight the power of recourse for \omm. Our first result is for general metric spaces:

\begin{theorem} 
	\label{thm:general}
	There is an efficient deterministic $2 \log k$-competitive algorithm with $\log k$-per-client recourse for \ommr on general metrics.
\end{theorem}
The above result is in contrast with the $(2k-1)$ lower bound for deterministic algorithms without recourse. Our algorithm uses recourse to mimic the output of a batched version of the classical \permutation algorithm for \omm, thereby highlighting the robustness of \permutation. \Cref{prop:batch-permutation-general} generalizes the above to give a cost-recourse trade-off. The guarantees given above are tight for our algorithm. We also prove a lower bound result:

\begin{theorem}
	No deterministic algorithm for \ommr with per-client recourse at most an absolute constant $C$ can have a competitive ratio $o(\log k)$.
\end{theorem}

%Our lower bound proof uses the star metric, and is the same instance as the hard instance for \permutation without recourse.  

For the line metric, we present a special-purpose algorithm that significantly improves on \Cref{thm:general}: %show that bounded recourse can also help in designing better algorithms for \omm on line metrics.

\begin{theorem} 
	\label{thm:line}
	There is a deterministic $3$-competitive algorithm with $\Ocal(\log k)$-amortized recourse for \ommlr.
\end{theorem}

Our algorithm is again based on \permutation. In a nutshell, when a new client $c_t$ arrives, we determine the free server $s_t$ which \permutation will match $c_t$ with. On a line metric, we can view this matching $(c_t,s_t)$ as a directed arc from $c_t$ to $s_t$ with cost exactly equal to the length of the arc. Noting that such a matching may be sub-optimal only due to the presence of overlapping forward and backward arcs, our algorithm tries to \emph{cancel} overlapping arcs using an uncrossing type of re-matching. However, blindly re-matching overlapping arcs leads to a large recourse, and we need to carefully determine the sequence in which we uncross to ensure a balance between competitive ratio and recourse. 
Towards this, we formulate and analyze an \emph{asymmetric} version of \permutation with canceling, called \farthestserver, which we believe is the key contribution of our work. 
A novel feature of our analysis is that we consider \emph{two different algorithms} which produce matchings of equal cost, and we use each of them to bound the cost and recourse respectively.

Finally, we turn our attention to another limitation of the classical \omm problem -- due to the irrevocable nature of assignments, the competitive ratio would be unbounded when both clients and/or servers can arrive or depart the system. Hence, the classical model only considers the setting when all servers are known ahead of time and clients arrive in an online manner. We show that by allowing recourse, we can handle arrivals and departures of clients and servers.

\begin{theorem} 
\label{thm:fully-online}
	There is an efficient randomized $\Ocal(\log n)$-competitive algorithm with $\Ocal(\log \Delta)$ amortized recourse for \ommr on general metrics when clients and servers can arrive and depart.
\end{theorem}

\subsection{Related Work}
To the best of our knowledge, the only work which considers recourse for online min-cost matching is the recent work \cite{matuschke2018maintaining} where the authors consider a \emph{two-stage} version of the uni-chromatic problem (where there is no distinction between servers and clients): In the first stage, a perfect matching between $2n$ given nodes must be selected; in the second stage $2k$ new nodes are introduced. The goal is to produce $\alpha$-competitive matchings at the end of both stages, and such that the number of edges removed from the first stage matching is at most $\beta k$. The authors show that $\alpha=3,\beta=1$ and $\alpha=10, \beta=2$ are possible when $k$ is known or unknown, respectively. Our results can be seen as a multi-stage generalization of this two-stage model, although the two models are slightly different in terms of the distinction between servers and clients.

 A related model which has received much attention recently, and which captures a different kind of flexibility in online matching, is that of \emph{matching with delays}~\cite{emek2016online,bienkowski2017match}: here, the requests do not have to be matched at the time of their arrival, but accrue a delay penalty until the algorithm matches it. The algorithm must minimize the total matching cost plus total delay penalty. The current best known randomized algorithms are  $\Ocal(\log n)$ competitive~\cite{ashlagi2017min}, which also shows a lower bound of $\Omega \left( \frac{\log n}{ \log \log n} \right)$. The best known deterministic algorithms are $\Ocal(k^{0.59})$-competitive \cite{azar2018deterministic}. Finally, another class of beyond-worst case models are stochastic models, such as {\it i.i.d.} and {\it random order} settings. The majority of work  in this vein has been done in the maximization objective rather than cost minimization (see e.g., \cite{goel2008online, devanur2012asymptotically, brubach2016new} and references therein). 
For \omm, \cite{raghvendra2016robust} gave a deterministic algorithm that is simultaneously $\Ocal(\log k)$-competitive in the random order model and  $(2k-1)$-competitive in worst case. Recently, \cite{gupta2019stochastic} show $\Ocal((\log \log \log k)^2)$-competitive algorithms in the known {\it i.i.d.} model.

Online algorithms with recourse have been studied in various other settings such as scheduling and set cover. We refer the reader to \cite{GKS14,GKKP17,FFGGKRW18} and the references therein, for online algorithms which make use of a small amount of recourse to get improved competitive ratio. 

% [TODO] Recourse for other problems, online service with delay, Online matching with maximization, Stochastic models (other ways to avoid badness in competitive ratio)

%Recognizing the need for algorithms with favorable performance in practical settings, the study of online matching has been extended to {\it i.i.d.} setting (the request locations are assumed to be sampled from a known or unknown underlying distribution on the metric) and {\it random order} (an adversarially generated set of locations is presented to the online algorithms in a uniformly sampled permutation). Bulk of the work in this vein has been done in the profit maximization rather than cost minimization setting motivated by online ad auctions (e.g., random order  \cite{goel2008online, karande2011online, mahdian2011online}, unknown i.i.d. \cite{devanur2012asymptotically, mirrokni2012simultaneous}, known i.i.d. \cite{bahmani2010improved, brubach2016new, feldman2009online}, and stochastic with noise  \cite{esfandiari2015online}). 
%For metric min-cost matching, \cite{raghvendra2016robust} gave a deterministic algorithm that is $\Ocal(\log k)$-competitive in the random order model, and simultaneously $(2k-1)$-competitive in the adversarial model. Very recently, \cite{gupta2019stochastic} propose a $\Ocal((\log \log \log k)^2)$-competitive algorithm in the known {\it i.i.d.} model of request arrivals. \cite{dehghani2017stochastic} consider the $k$-server problem in the known {\it i.i.d.} model -- they propose a $3$-competitive algorithm for line and circle metric and $\Ocal(\log n)$ for general metric. 

\subsection{Outline}
In~\Cref{sec:prelims} we begin with some useful notation.
Then in~\Cref{sec:permutation}, we describe the  \permutation algorithm~\cite{KP93,khuller1994line} which is crucial to all of our algorithms. In Section~\ref{sec:general-metrics}, we present our algorithm for general metrics and prove~\Cref{thm:general}. Next we turn our attention to the line metric in~\Cref{sec:line} and prove~\Cref{thm:line}. Finally we consider the fully dynamic setting in~\Cref{sec:fully-online} and prove~\Cref{thm:fully-online}.

\section{Preliminaries}
\label{sec:prelims}
For most of the paper (except the fully dynamic setting), we consider the setting where the servers $\Scal$ are known up front. The clients arrive online, and we denote by $C_t = (c_1, \ldots, c_t)$ the set of first $t$ clients. 
An optimal matching between $C_t$ and $\Scal$ is denoted by $\Mcal^*_t$, and similarly, the matching maintained by the algorithm between $C_t$ and $\Scal$ will be denoted by $\Mcal_t$. We denote by $\OPT_t$, the cost of the optimal matching $\Mcal^*_t$. 
For any matching $\Mcal$, we use $\Mcal(c)$ and $\Mcal(C)$ to denote the server and the set of servers matched to the client $c$ and the set of clients $C$, respectively. We define $\Mcal(s)$ and $\Mcal(S)$ similarly.

%and by $M_t$ the matching of the online algorithm after the $t$th customer has been matched. We denote the set of servers matched in $M_t$ by $S_t$. $\Mcal_t$ denotes the optimal matching of $C_t$ to $S$, and $\Scal_t$ the set of servers matched in $\Mcal_t$. The final matchings are denoted by $M \equiv M_n$ and $\Mcal \equiv \Mcal_n$. We will find it convenient to view a (partial) matching $M_t$ as a bipartite graph between the $C_t \cup S$ where an edge $(c,s)$ exists in the bipartite graph if customer $c$ is matched to server $s\in S$ under matching $M_t$.
%For a (partial) matching, $M'$, $\cost(M')$ denotes the sum of the distances of the matched client-server pairs in $M'$. We abbreviate $\OPT_t = \cost(\Mcal_t)$ and $\OPT = \cost(\Mcal)$. The aspect ratio of a metric space $(\Xcal, d)$ is defined as $\Delta \doteq \max_{ x,y \in \Xcal} d(x,y) / \min_{ x \neq y \in \Xcal } d(x,y)$.

\section{The \permutation algorithm} \label{sec:permutation}
As mentioned in~\Cref{sec:intro}, \cite{KP93} and \cite{khuller1994line} independently proposed a $(2k-1)$-competitive algorithm \permutation for \omm. Since our algorithms build extensively on this algorithm, we first  describe \permutation and its key properties. 
The algorithm maintains two matchings: the current online matching $\Mcal_t$, and the optimal offline matching  $\mathcal{M}^*_t$ of the clients $C_t$ that have arrived so far. The main observation behind the algorithm is that, when a new client $c_{t+1}$ arrives, there exists an optimal matching of $C_{t+1}$ to $\Scal$ which uses exactly the servers used in $\Mcal^*_t$ plus one extra server. \permutation simply identifies the extra server $s_{t+1}$ and matches $c_{t+1}$ with $s_{t+1}$. This property can be formalized as follows.
%The next Lemma proves a slightly more general version of this property of optimal matchings exploited by \permutation (we defer all proofs to Appendix~\ref{sec:permutation-proofs}).

\begin{lemma} \label{lem:offline-monotone}
There exists a sequence of optimal matchings $\Mcal^*_1, \ldots, \Mcal^*_k$ matching client sets $C_1 \subseteq C_2 \subseteq \cdots \subseteq C_k$ to $\Scal$ such that the sets of servers used in these matchings, $S^*_i := \Mcal^*_i(C_i)$ are nested, i.e., $S^*_1 \subseteq S^*_2 \subseteq \cdots \subseteq S^*_k$. 
\end{lemma}
\begin{proof}
	Consider two sets of clients $C_{t_1} \subset C_{t_2}$. Let $\Mcal_{t_1}$ denote any optimal min-cost matching of all the clients in $C_{t_1}$ to $\Scal$, and let $S_{t_1} \subseteq \Scal$ denote the servers matched in $\Mcal_{t_1}$. Then, it suffices to prove that there exists an optimal min-cost matching $\Mcal_{t_2}$ of all the clients in $C_{t_2}$ to $\Scal$, such that the set of matched servers $S_{t_2} \subseteq \Scal$ in $\Mcal_{t_2}$ is an extension of $S_{t_1}$, i.e., $S_{t_1} \subseteq S_{t_2}$ with $|S_{t_2} \setminus S_{t_1}| = |C_{t_2} \setminus C_{t_1}|$. 
	
	Let $\Mcal_{t_2}$ be a minimum cost matching between the clients $C_{t_2}$ and the servers $\Scal$ which has the smallest $| S_{t_2} \setminus S_{t_1}|$ with $S_{t_2} := \Mcal_{t_2}(C_{t_2}), S_{t_1} = M_{t_1}(C_{t_1})$. Consider the graph $\Mcal_{t_1} \cup \Mcal_{t_2}$. Since every vertex in this graph has degree at most $2$, this is a union of cycles and paths. Moreover, the clients in $C_{t_2} \setminus C_{t_1}$ have degree exactly equal to 1, and the clients in $C_{t_1}$ have degree 2. Servers which are part of cycles in $\Mcal_{t_1} \cup \Mcal_{t_2}$ are part of both $\Mcal_{t_1}$ and $\Mcal_{t_2}$  and therefore can be ignored for counting $|S_{t_2} \setminus S_{t_1}|$. We have two types of paths:
	\begin{enumerate}
		\item Paths which begin with a client, which must necessarily be in $C_{t_2} \setminus C_{t_1}$: Since the first edge of this path belongs to $\Mcal_{t_2}$ and the edges alternate between $\Mcal_{t_2}$ and $\Mcal_{t_1}$ while nodes alternate between client and server, we must have that the path ends in a server in $S_{t_2}\setminus S_{t_1}$. On this path, there is exactly one client in $C_{t_2} \setminus C_{t_1}$, and exactly one server in $S_{t_2} \setminus S_{t_1}$.
		\item Paths which begin with a server in $S_{t_1} \setminus S_{t_2}$: The first edge of this path is in $\Mcal_{t_1}$, and the edges alternate between $\Mcal_{t_1}$ and $\Mcal_{t_2}$ while nodes alternate between server and client. Such a path can not end in a client because this would then have to be a client in $C_{t_1}$ which have degree 2. Therefore this path ends in a server in $S_{t_2}\setminus S_{t_1}$, and further has an even number of edges. Therefore, all clients on this path are in $C_{t_1}$, and the $\Mcal_{t_1}$ edges and the $\Mcal_{t_2}$ edges give two different matchings for these clients. Since we assume that both $\Mcal_{t_1}$ and $\Mcal_{t_2}$ are optimal, these matchings must be of the same cost. But then, by switching the matches for the clients on this path from those in $\Mcal_{t_2}$ to $\Mcal_{t_1}$ we obtain another min cost matching $\Mcal_{t_2}'$ with server set $S_{t_2}'$ satisfying $|S_{t_2}' \setminus S_{t_1}| = |S_{t_2} \setminus S_{t_1}| -1 $ violating the assumption that $\Mcal_{t_2}$ has the largest overlap with $\Mcal_{t_1}$.
		\item Paths which begin with server in $S_{t_2} \setminus S_{t_1}$: Such paths must be one of the two above types, and therefore do not require special analysis.
	\end{enumerate}
	In summary, assuming the optimality of $\Mcal_{t_2}$ as the min cost matching of $C_{t_2}$ with the largest overlap with $\Mcal_{t_1}$ in terms of server set, every client in $C_{t_2} \setminus C_{t_1}$ contributes exactly one server to $S_{t_2}\setminus S_{t_1}$ given by the server at the end of the augmenting path started by the client in $\Mcal_{t_1} \cup \Mcal_{t_2}$.
\end{proof}

\begin{definition}[Server-optimal matching] \label{def:server-optimal-matching}
At time $t$, a matching $\Mcal_t$ of client-set $C_t$ is said to be \emph{server-optimal} if it uses the same servers as $\Mcal^*_t$, i.e.,  $\Mcal^*_t(C_t) = \Mcal_t(C_t)$. 
\end{definition}

\begin{algorithm}[t]
	\caption{\permutation (metric $(\Xcal, d)$, server-optimal matching $\Mcal_{t-1}$ for client set $C_{t-1}$)}
\label{alg:perm}
		\begin{algorithmic}[1]
		\For {new batch of clients $C_{\rm{cur}} = C_{t + \ell} \setminus C_{t-1} = \{c_{t}, c_{t+1}, \ldots, c_{t+\ell}\}$ that arrives}
		\LState let $\Mcal^*_{t-1}$ and $\Mcal^*_{t+\ell}$ be  optimal matchings for $C_{t-1}$ and $C_{t+\ell}$ from~\Cref{lem:offline-monotone}. 
		\LState let $S_{\rm{cur}} = S^*_{t+\ell} \setminus S^*_{t-1}$ denote the set of $\ell+1$ servers matched in $\Mcal^*_{t+\ell}$ but not in $\Mcal^*_{t-1}$ \label{alg1:step3}
		\LState let $\Mcal_{{\rm cur}}$ denote the minimum cost matching between $C_{\rm{cur}}$ and $S_{\rm{cur}}$ \label{alg1:step4}
		\LState augment $\Mcal_{t-1}$ using $\Mcal_{{\rm cur}}$ to obtain the new matching $\Mcal_{t + \ell}$  \label{alg1:step5}
		\EndFor
	\end{algorithmic}
	%\end{center}
\end{algorithm}

\begin{proposition} \label{obs:offline-used}
\permutation always maintains a server-optimal matching.
\end{proposition}

Algorithm~\ref{alg:perm} gives a more general version of \permutation which we will use later, where clients arrive in batches. Lemma~\ref{lem:new-batch} gives a bound on the increase in the cost of the matching maintained by \permutation after each batch of clients, culminating in Theorem~\ref{thm:permutation}. 
%We defer all proofs to Appendix~\ref{sec:permutation-proofs}.

\begin{lemma} \label{lem:new-batch}
After the arrival of a batch of clients $C_{t + \ell} \setminus C_{t-1}$, the cost of the matching $\Mcal_{{\rm cur}}$ computed in~\Cref{alg1:step4} is at most $2 \OPT_{t+\ell}$.
\end{lemma}
\begin{proof}
	Let $\Mcal^*_{t+\ell}$ denote an optimal matching of $\Ccal_{t+\ell}$ using the servers $\Scal_{t+\ell}^*$, and $\Mcal^*_{t-1}$ denote an optimal matching of $\Ccal_{t-1}$ using $\Scal_{t-1}^*$. Consider the graph $\Mcal^*_{t+\ell} \cup \Mcal^*_{t-1}$. As in the proof of \Cref{lem:offline-monotone}, this graph contains cycles, or paths which start from a client in $\Ccal_{\rm cur}$ and ends at a server in $\Scal_{\rm cur}$. Matching the clients and servers at the end points of these augmenting paths defines one way to match the clients in $\Ccal_{\rm cur}$ to servers in $\Scal_{\rm cur}$, and because of the triangle inequality, the total cost of this matching of $\Ccal_{\rm cur}$ to $\Scal_{\rm cur}$ is at most the total cost of the augmenting paths, which is bounded by $\cost(\Mcal^*_{t+\ell})+\cost(\Mcal^*_{t-1})$. Finally, since $\Mcal_{\rm cur}$ is the minimum cost matching of $\Ccal_{\rm cur}$ to $\Scal_{\rm cur}$, we get:
	\[ \cost(\Mcal_{\rm cur}) \leq \cost(\Mcal^*_{t+\ell}) + \cost(\Mcal^*_{t-1}) \leq 2 \cost(\Mcal^*_{t+\ell}) = 2 \OPT_{t+\ell}.\]
	The last inequality follows since the cost of the optimal matching is monotonically non-decreasing in time.
\end{proof}

\begin{theorem}(Theorem $2.4$ in~\cite{KP93})
\label{thm:permutation}
\Cref{alg:perm} is $(2m-1)$-competitive for online weighted matching if the requests arrive in $m$ batches. \end{theorem}
\begin{proof}
	Let the size of the $m$ batched me $\ell_1, \ldots, \ell_m$,  $L_j := \ell_1 + \ell_2 + \cdots + \ell_j$, the matching produced by Algorithm~\ref{alg:perm} after the $j$th batch be denoted by $\Mcal_{L_j}$ and the optimal matching for clients $\Ccal_{L_j}$ be $\Mcal^*_{L_j}$. Then, using \Cref{lem:new-batch},
	\begin{align*}
	\cost(\Mcal_{L_m}) &= \cost(\Mcal_{L_1}) + \sum_{j=2}^m \cost(\Mcal_{L_{j}}) - \cost(\Mcal_{L_{j-1}}) \\
	& \leq \cost(\Mcal^*_{L_1}) + \sum_{j=2}^m 2 \cdot \cost(\Mcal^*_{L_j}) \\
	& \leq (2m-1) \cost(\Mcal^*_{L_m}) = (2m-1) \OPT_{L_m}.
	\end{align*}
\end{proof}

\section{Online matching with recourse for general metrics} \label{sec:general-metrics}

In this section, we present our $(\Ocal(\log k), \Ocal(\log k))$-competitive algorithm for arbitrary metrics. 
Indeed, \Cref{thm:permutation} says that in order to minimize competitive ratio, it is best to feed the input to \permutation in as few batches as possible. However, this idea is in contradiction with the rule that the online algorithm must match clients immediately on arrival. One way of balancing the two goals is to actually run \permutation incrementally on each client arrival, but when a group of clients has arrived, we un-match the current matching for this group and re-introduce all these clients as \emph{a single batch}, thereby exploiting the power of recourse. As an example, assume that we create $\sqrt{k}$ batches of $\sqrt{k}$ clients, with the $j$th batch consisting of clients ${\rm Batch}_j = \{(j-1)\sqrt{k}+1, \ldots, j\sqrt{k}\}$. While clients in batch $j$ arrive, we first run vanilla \permutation, matching the new client to the server added to the optimal matching. After the $j\sqrt{k}$th client arrives, we un-match all clients in ${\rm Batch}_j$ and re-introduce them as one single batch. The amortized recourse of this algorithm is 1, and moreover, the matching at any time $t$ may be viewed as the output of running \permutation with  $\floor {\frac{t}{\sqrt{k}}} \leq \sqrt{k}$ batches of $\sqrt{k}$ clients and $\left(t - \sqrt{k}\floor {\frac{t}{\sqrt{k}}} \right) \leq \sqrt{k}$ batches of $1$ client. 

To get a smaller competitive ratio at the expense of slightly higher recourse, we employ the following natural extension: imagine we run $O(\log k)$-parallel runs of \permutation, with the $i^{th}$ run operating in batches of size $2^i$. Then we can bound the competitive ratio by $2 \log k$ if we can ensure that the matching at time $t$ is simply the combination of various matchings based on the binary decomposition of $t$. Indeed, the following algorithm precisely achieves this property for all $1 \leq t \leq k$, while just using a per-client recourse of $\log k$.

\begin{figure}
\centering
\begin{subfigure}[b]{0.5\textwidth}
  \centering
  \includegraphics[width=0.9\linewidth]{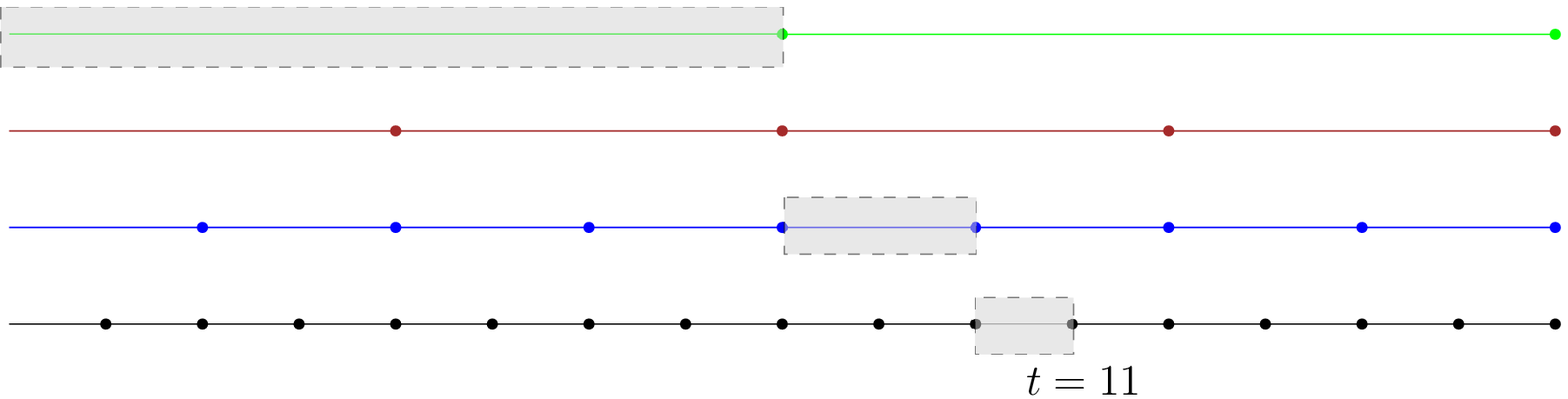}
  \caption{$\Mcal_{11}$ is the union of three blocks of length $8$, $2$ and $1$.}
  \label{fig:batchperm11}
\end{subfigure}%
\begin{subfigure}[b]{0.5\textwidth}
  \centering
  \includegraphics[width=0.9\linewidth]{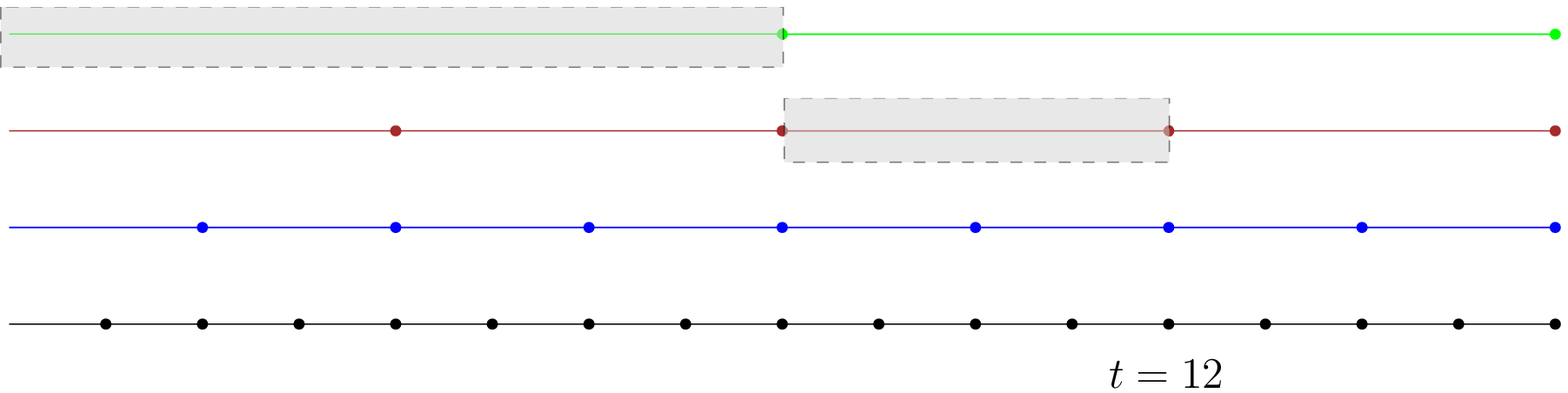}
  \caption{How $\Mcal_{11}$ is changed to $\Mcal_{12}$.}
  \label{fig:batchperm12}
\end{subfigure}
\vspace{0.01cm}
\caption{Illustration of \batchperm}
\label{fig:batchperm}
\end{figure}

\begin{algorithm}
	\caption{\batchperm (metric $(X,d)$ and server set $\Scal \subseteq X$)}
\label{alg:batch-perm}
		\begin{algorithmic}[1]
		\State initialize matching $\Mcal_0 = \emptyset$
		\For {each new client $c_t$ that arrives at time-step $t$}  \label{alg2:step1}
		\State let $i(t) = \arg \max_{i}$ s.t $t$ is divisible by $2^i$ \label{alg2:step2} 
		\State un-match the latest $2^{i(t)}$ clients $\{c_{t - 2^{i(t)}+1}, \ldots, c_{t}  \}$ and revert back to the matching $\Mcal_{t - 2^{i(t)}}$ \label{alg2:step3}
		\State introduce a block of clients $\{c_{t - 2^{i(t)}+1}, \ldots, c_{t}  \}$ to ~\Cref{alg:perm} with the current matching being $\Mcal_{t - 2^{i(t)}}$ and update $\Mcal_t$ to be the resulting matching for all the clients $C_t$ \label{alg2:step4}
		\EndFor
	\end{algorithmic}
	%\end{center}
\end{algorithm}

\begin{theorem}
\label{thm:batch-permutation}
At any time $t$, the total recourse of the Algorithm~\ref{alg:batch-perm} is bounded by $\Ocal(t \log t)$. Furthermore, the cost of the matching $\Mcal_t$ is at most $\Ocal(\log t)$ times the optimal offline matching  $\Mcal^*_t$.
\end{theorem}

\begin{proof}[Proof Sketch]
At any time $t$, we view our algorithm as simulating the \permutation algorithm for a certain batch sequence. Indeed, note, the solution maintained in $\Mcal_t$ is exactly what \permutation maintains when fed $O(\log t)$ batches of consecutive clients corresponding to the different powers-of-two $2^{i-1}$ (in decreasing order) such that the $i^{th}$ bit from right in the binary representation of $t$ is  $1$. \Cref{thm:permutation} then bounds the cost. The recourse is bounded since any client is involved in a re-matching of size $2^i$ at most once for all $i$.
\end{proof}

The next proposition generalizes the result in Theorem~\ref{thm:batch-permutation} to give a trade-off between the cost and recourse metrics. Proposition~\ref{prop:batch-permutation-tight} proves that the analysis is tight for \batchperm. That is, there are instances where \batchperm indeed achieves the cost and recourse mentioned. The proof of Proposition~\ref{prop:batch-permutation-tight} appears in Appendix~\ref{sec:general-metrics-proofs}.

\begin{proposition}
\label{prop:batch-permutation-general}
Algorithm~\ref{alg:batch-perm} with the constant $2$ replaced by $d$, gives an $((d-1) \log_{d}k , \log_{d} k)$-competitive algorithm. In particular, for any $d=\Ocal(1)$ we get $(\Ocal(\log k), \Ocal(\log k))$-competitive, and for $d=k^{\alpha}$ ($\alpha \leq 1$), we get an $(\tilde{\Ocal}(k^\alpha), 1+1/\alpha)$-competitive algorithm.
\end{proposition}

\begin{proposition}
\label{prop:batch-permutation-tight}
The cost-recourse tradeoff of Proposition~\ref{prop:batch-permutation-general} is tight for Algorithm~\ref{alg:batch-perm}.
\end{proposition}

\section{Online Matching on the Line metric}
\label{sec:line}

In this section we focus on the special case of a line metric. That is, for all points $x \in \Xcal$, we associate a location $\ell:\Xcal \to \RR$, and $d(x,y) = |\ell(x) - \ell(y)|$. 
Furthermore, we assume that all the clients and servers are in distinct locations on the line.  
Before we describe our algorithm, we first explain some structural properties about optimal matchings on the line metric space.

\begin{definition}[Forward and Backward Arcs] \label{def:arcs}
Suppose a client $c$ is matched to server $s$ in some matching $\Mcal$. Then, we interchangeably represent this edge $(c,s)$ as an \emph{arc from $c$ to $s$}. Moreover, it is said to be a \emph{forward arc} $ \overrightarrow{c s}$ if $\ell(c) \leq \ell(s)$ and a \emph{backward arc} $ \overleftarrow{s c}$ otherwise. 
\end{definition}

The following observation says that opposite arcs in an optimal matching are \emph{non-overlapping}.

\begin{observation} \label{obs:no-crossing-arcs-opt}
	Consider a set of clients $C$ and set of servers $S$ with $|C| = |S|$. Then, any matching $\Mcal$ between $C$ and $S$ is optimal if and only if, for every pair of forward arc $\overrightarrow{c_1 s_1} \in \Mcal$ and backward arc $\overleftarrow{s_2 c_2} \in \Mcal$, the intervals $[\ell(c_1), \ell(s_1)]$ and $[\ell(s_2) , \ell(c_2)]$ are disjoint.
\end{observation}

\begin{figure}
	\centering
	\begin{subfigure}[t]{0.32\textwidth}
		\centering
		\includegraphics[width=0.95\linewidth]{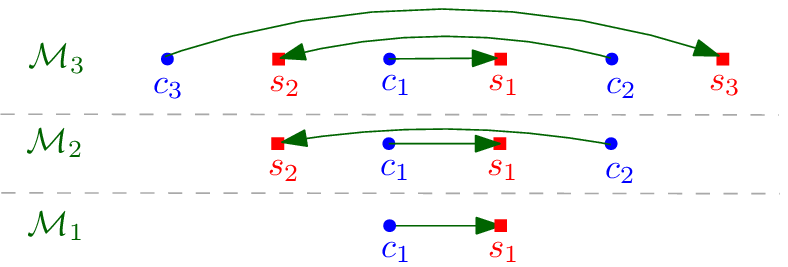}
		\caption{\permutation without re-match}
		\label{fig:line1a}
	\end{subfigure}%
	\begin{subfigure}[t]{0.32\textwidth}
		\centering
		\includegraphics[width=0.95\linewidth]{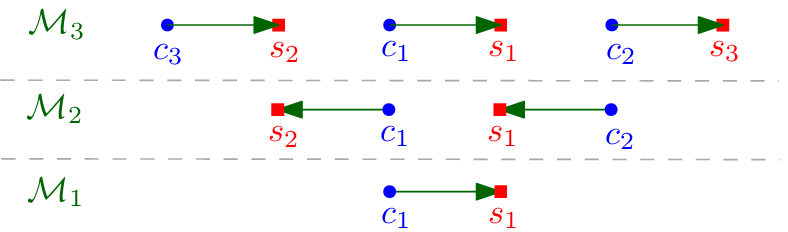}
		\caption{Symmetric  \permutation with re-match}
		\label{fig:line1b}
	\end{subfigure}
	\begin{subfigure}[t]{0.32\textwidth}
	\centering
	\includegraphics[width=0.95\linewidth]{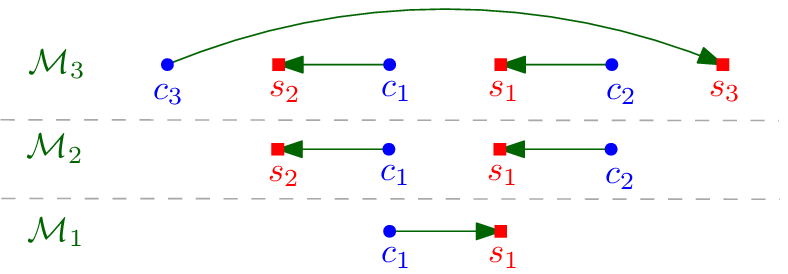}
	\caption{\emph{Asymmetric} \permutation with re-match}
	\label{fig:line1c}
\end{subfigure}
	\caption{Illustrative examples of \ommlr}
	\label{fig:line1}
\end{figure}

Based on~\Cref{obs:no-crossing-arcs-opt}, a natural approach for \ommlr would be to run \permutation, which by itself can have cost as bad as $\Omega(k) \OPT$ (see \Cref{fig:line1a}), and remove all overlapping arcs by re-matching suitably. Indeed, if $\overrightarrow{c_1 s_1}$ and $\overleftarrow{s_2 c_2}$ overlap, then we could re-match $c_2$ to $s_1$ and $c_1$ to $s_2$ to make this pair non-overlapping. By doing this repeatedly, we can make the matching non-overlapping. Moreover, since \permutation is server-optimal, we would then have an optimal solution at the end! Unfortunately though, it is easy to construct examples where this procedure would result in as many as $\Omega(k)$ re-matches per client (see~\Cref{fig:line1b}).  To summarize, \permutation does no recourse but has large competitive-ratio, and always re-matching overlapping arcs output by \permutation yields optimal solutions with large recourse. 

Our idea, perhaps natural in hindsight, is to then balance the cost and recourse by re-matching overlapping pairs \emph{asymmetrically}. Indeed, when \permutation adds a new forward arc $\overrightarrow{c s}$, our algorithm also does the same, and \emph{only} tries to cancel/re-match overlapping intervals when \permutation adds a backward arc $\overleftarrow{s c}$ (see~\Cref{fig:line1c} for an example where $\Mcal_2$ has undergone a re-matching, while $\Mcal_3$ has not). While this is unambiguous for the example in~\Cref{fig:line1c}, in general, there could be multiple ways of re-matching overlapping arcs. Indeed, our final algorithm, called \farthestserver, identifies one such way of re-matching for which the recourse is $O(\log k)$. However, the cost analysis of \farthestserver seems tricky.
Towards this, we approach the problem indirectly and introduce \emph{another algorithm}, called \recursivecancel, and a) show that the cost of \recursivecancel and \farthestserver are identical, and b) bound the cost of \recursivecancel by $3 \, \OPT$.   

We remark that the idea of asymmetrically re-matching only backward arcs crucially uses the fact that \permutation is server-optimal (\Cref{def:server-optimal-matching}), and this method does not yield $\Ocal(1)$ approximation for arbitrary sets of clients and servers. This is in contrast to re-matching all overlapping forward and backward arcs, which computes the optimal matching for any set of client-server sets. To see why, consider the following simple example: suppose $\ell(c_1) = 1$, and $\ell(s_1) = 0$. Now, imagine a new client arrives with $\ell(c_2) = 0$ and suppose $s_2$ was determined to be at location $\ell(s_2) = 1$. The optimal matching for this clearly has cost $0$, while the asymmetric algorithm would keep both the backward arc (since there are no cancellations possible at this time), and the forward arc (since the asymmetric algorithm does nothing when forward arcs are added). However, what saves us, is the fact that such situations cannot arise when using \permutation, which is server-optimal at all times. Indeed, if \permutation adds a forward arc $(c_2,s_2)$, we can conclude that $\ell(s_2) \geq 2$ due to server optimality at time $1$. Hence, we can bound the cost of $\Mcal_2$ by $3 \, \OPT$.
%This property, formally captured by~\Cref{obs:offline-used}, will be crucial to our proof. 

\medskip \noindent {\bf Intervals and Discrepancy.} 
At any time $t$, both our algorithms will use exactly the same set of servers $S_t$ as used by \permutation, which satisfies $S_t = \Mcal^*_t(C_t)$ from server-optimality.
We divide the line into \emph{intervals} corresponding to open intervals between two consecutive points in $C_t \cup S_t$ in the metric space. In the analysis, we will also  label every interval of every arc as either \emph{redundant} or \emph{non-redundant}, and so we sometimes abuse notation and refer to intervals in an arc-specific way as ``interval of an arc''. 
For an interval $I = (l, r)$, we denote by $\disc_t(I) =  \abs{S_t \cap (-\infty, l]} - \abs{C_t \cap (-\infty, l]}$ to be the \emph{discrepancy} of $I$ at time $t$. In words, it is the excess number of servers over clients to the left of $I$. Indeed, if $\disc_t(I)$ is negative (resp. positive), there there will be $\disc_t(I)$ forward (resp. backward) arcs crossing $I$ in an optimal matching between $C_t$ and $S_t$. We also keep track of the following quantities for the algorithm's matching: for interval $I = (l, r)$, let $n^f_t(I)$ (resp. $n^b_t(I)$) denote the number of forward (resp. backward) arcs crossing $I$ at time $t$. 

\medskip \noindent {\bf Maximally-Canceling Algorithms.} 
Before we formally describe them, we mention a nice property about \farthestserver and \recursivecancel. Both algorithms can be described by the following common framework: (a) when a new client $c_t$ arrives, run \permutation and let $s_t$ be the new server \permutation brings into the system. (b) if the matching $(c_t,s_t)$ is a forward arc, i.e., $\ell(c_t) \leq \ell(s_t)$, then simply add it to $\Mcal_{t-1}$ to get $\Mcal_t$; otherwise, consider all forward arcs in $\Mcal_{t-1}$ with \emph{overlap} with the backward arc $(c_t, s_t)$, and \emph{maximally cancel} overlapping intervals by re-matching. That is, after the re-matching is done,  if you consider any interval $I$ in $[\ell(s_t), \ell(c_t)]$ such that $n^f_{t-1}(I) > 0$, it will hold that $n^f_t(I) = n^f_{t-1}(I) - 1$ and $n^b_t(I) = n^b_{t-1}(I)$, and
if interval $I$ in $[\ell(s_t), \ell(c_t)]$ has  $n^f_{t-1}(I) = 0$, then it holds that $n^f_t(I) = 0$ and $n^b_t(I) = n^b_{t-1}(I) + 1$. This will in fact establish that the two algorithms have the same cost, since the cost of a matching can be expressed as $\sum_{I} |I| \left( n^f_t(I) + n^b_t(I) \right)$ over all intervals $I$, with $|I|$ denoting the length.

\subsection{Algorithm \recursivecancel For Bounding Cost} \label{sec:linecost}
We now present our algorithm \recursivecancel (\Cref{alg:recursivecancel}) and bound its cost. In~\Cref{sec:linerecourse} we present our actual algorithm \farthestserver and bound its recourse. Since both algorithms will be maximally-canceling, we can bound the cost of \farthestserver as well, thereby proving~\Cref{thm:line}.

\newcommand{\mnew}{\Mcal_f^{\rm new}}
\newcommand{\mold}{\Mcal_f^{\rm old}}

\begin{algorithm}
	\caption{Algorithm \recursivecancel}
\label{alg:recursivecancel}
\begin{algorithmic}[1]
    \LState set $\Mcal_0 = \emptyset$
	\For {each client $c_t$ arriving at time $t \geq 1$}
	\LState let $s_t$ be the server \permutation matches $c_t$ to, and let $a := (c_t,s_t)$
\If {$a$ is a forward arc}
	\LState {$\Mcal_{t} = \Mcal_{t-1} \cup \{a\}$} \label{alg3:step4}
		\Else \Comment{$(c_t, s_t)$ is a backward arc}
	\While {there exists forward arcs in $\Mcal_{t-1}$ which overlaps with $a$} \label{alg3:step7} 	\Comment {$a$ is the current backward arc}
	\LState let $a := (c,s)$ \label{alg3:step8}
	\LState let $(c',s')$ be a forward arc overlapping with $a$ with the \emph{rightmost} server $s'$ \label{alg3:step9}
	\LState $\Mcal_{t-1} = \Mcal_{t-1} \setminus \{(c',s') \} \cup \{ (c,s') \}$  \label{alg3:step10}
	\LState set $a := (c',s)$ \label{alg3:step11} \Comment{From~\Cref{lem:no-middle-server}, $a$ will be a backward arc for loop recursion} \label{alg3:step12}
	\EndWhile\label{alg3:step13}
	\LState $\Mcal_t = \Mcal_{t-1} \cup \{a\}$  \Comment{The final $a$ has no overlapping forward arcs, and is added to $\Mcal_t$}
	\EndIf
	\EndFor
\end{algorithmic}
\end{algorithm}

We start with a couple of simple yet crucial lemmas. 

\begin{lemma} \label{lem:no-middle-server}
For any arc $(c,s) \in \Mcal_t$, there is no unmatched server available at location $x \in [\ell(c),\ell(s)]$. 
\end{lemma}
\begin{proof}
	We will first prove that if we execute the \permutation algorithm on line, there are no free servers inside any arc. 
	Recall that \permutation maintains an offline optimal matching $\mathcal{M}_t^*$ at time $t$, and when a client $c_t$ arrives, we pair it with the server that is present in $\mathcal{M}_t^* \setminus \mathcal{M}_{t-1}^*$. 
	In fact, the symmetric difference of $\mathcal{M}_t^*$ and $\mathcal{M}_{t-1}^*$ is an augmenting path starting at $c_t$ and ending at $s_t$. 
	Let it be denoted by $P=c_t,s_{p_1},c_{p_1},\ldots,s_{p_m}=s_t$. The edges $(c_t,s_{p_1}),(c_{p_1},s_{p_2}),\ldots, (c_{p_{m-1}},s_{p_m})$ are the new edges, and the rest $(s_{p_1},c_{p_1}),(s_{p_2},c_{p_2}),\ldots, (s_{p_{m-1}},c_{p_{m-1}})$ are the old edges. 
	Recall that the cost of $\mathcal{M}_{t}^*$ is at least that of $\mathcal{M}_{t-1}^*$, and thus, in the augmenting path, the cost of new edges is at least that of the old edges.
	
	We claim a stronger property that in any suffix of the augmenting path, the cost of the new edges is at least that of old edges. 
	Consider a suffix $s_{p_i},c_{p_i},\ldots,s_{p_m}$. 
	If the cost of new edges is less than the old edges, we can change the old matching from $(s_{p_i},c_{p_i}),\ldots,(s_{p_{m-1}},c_{p_{m-1}})$ to $(c_{p_i},s_{p_{i+1}}),\ldots,(c_{p_{m-1}},s_{p_m})$ while keeping the rest of the edges intact to get a matching with cost less than $\mathcal{M}_{t-1}^*$, contradicting the fact that $\mathcal{M}_{t-1}^*$ is an optimal matching for the first $t-1$ clients. 
	Now, suppose that there is a free server $s'$ in between $c_t$ and $s_t$. 
	Consider the prefix of the augmenting path $P$ starting at $c_t$ and ending at $s'$. Let this prefix be denoted by $P'$. 
	Let $\mathcal{M}'_{t}$ be the matching obtained from $\mathcal{M}_{t-1}^*$ by augmenting with the new augmenting path $P'$. 
	Since the cost of new edges is at least that of old edges in any suffix of the original augmenting path, the difference between new edges and old edges in $P'$ is at most that of the original augmenting path $P$. 
	Thus, the cost of the matching $\mathcal{M}'_t$ is at most that of $\mathcal{M}_{t}^*$. 
	Furthermore, as we have assumed that the location of all the clients and servers are distinct, the cost of $\mathcal{M}'_t$ is strictly smaller than $\mathcal{M}_{t}^*$, contradicting the fact that $\mathcal{M}_{t}^*$ is an optimal matching of first $t$ clients. 
	This proves the claim that when we execute \permutation on the line metric, there are no free servers inside any arc.
	
	Using this, we will prove using induction on $t$ that after the arrival of $t$ clients, there are no free servers inside any arcs when executing \recursivecancel. 
	At $t=1$, we are adding an edge directly from \permutation, hence the claim follows trivially. 
	Consider the scenario when we are adding a client server pair $(c_t,s_t)$. 
	If the edge $(c_t,s_t)$ is a forward arc, we add it directly, and all other matches are unaffected. 
	As $(c_t,s_t)$ is added directly from \permutation, there are no free servers inside it, and thus proving the inductive claim. 
	Consider the case when $(c_t,s_t)$ is a backward arc. 
	As this edge is given by \permutation, there are no free servers inside the backward arc.
	This combined with the fact that the cancellation of the algorithm only adds segments inside $[\ell(s_t),\ell(c_t)]$ to arcs ensures that all the new arcs formed during recursive addition of $(c_t,s_t)$ don't contain any free servers. 
\end{proof}

\begin{lemma}
	\label{lem:shrink}
	Suppose the matched client of a server $s$ is changed from $c_1$ to $c_2$ during the course of \recursivecancel. Then, $\ell(c_2) \geq \ell(c_1)$. 
\end{lemma}
\begin{proof}
	Note that once a server is matched by a backward arc, it is not going to change its match from that point onward. Hence, we can assume that $c_1$ to $s$ is a forward arc. 
	Also note that in a single iteration of the algorithm, each server is rematched at most once in the recursive step. 
	
	Consider the iteration of ~\Cref{alg:recursivecancel} in which we rematch server $s$ from $c_1$ to $c_2$.  
	Let $(s',c')$ be the backward arc that was obtained from \permutation. During the recursive step of the iteration, $c_1,s$ intersected with backward arc $c_2,s'$, and thus the match of $s$ is changed from $c_1$ to $c_2$.
	This implies that $c_2$ is to the right of $c_1$ since otherwise, the two arcs would not have intersected in the first place.
\end{proof}

Note that the algorithm \recursivecancel is \emph{server-optimal} i.e. it uses the same set of servers as the optimal matching. 
In an interval $I$, the value $disc_t(I)= n_t^b(I)-n_t^f(I)$ is the same for all algorithms using the same set of servers as the optimal algorithm as it depends only on the set of servers used, not on the matching between clients and servers. 
Thus, informally speaking, algorithms that don't have good competitive ratio have both $n_t^b(I)$ and $n_t^f(I)$ high in some intervals due to which they pay extra cost. 
This motivates that in any interval, there are certain ``non-redundant'' arcs that optimal algorithm has to maintain as well, and additional ``redundant'' arcs which our algorithm has, but can be avoided by re-matching. 
More formally, we can define \emph{redundant} and \emph{non-redundant} arc intervals as follows:
\begin{definition}[Redundant and Non-Redundant arc intervals]
Suppose at time $t$, client $c_t$ is matched to server $s_t$ with a forward arc in~\cref{alg3:step4} of~\Cref{alg:recursivecancel}. 
		Then, an interval $I \in [\ell(c_t), \ell(s_t)]$ of this arc is said to be \emph{redundant} if $n^b_{t-1}(I) > n^f_{t-1}(I)$, and \emph{non-redundant} otherwise. Alternatively, if a new forward arc $(c,s')$ is added in the while loop~\cref{alg3:step10}, then an interval $I$ in the new arc simply is defined to be redundant if and only if the corresponding interval is redundant in arc $(c',s')$, where $c'$ is the client that $s'$ is matched to before $c$. Note that this definition makes sense because~\Cref{lem:shrink} implies that $c'$ is guaranteed to be on the left of $c$, so intervals of a new forward arc are indeed intervals of the previous forward arc.
\end{definition}

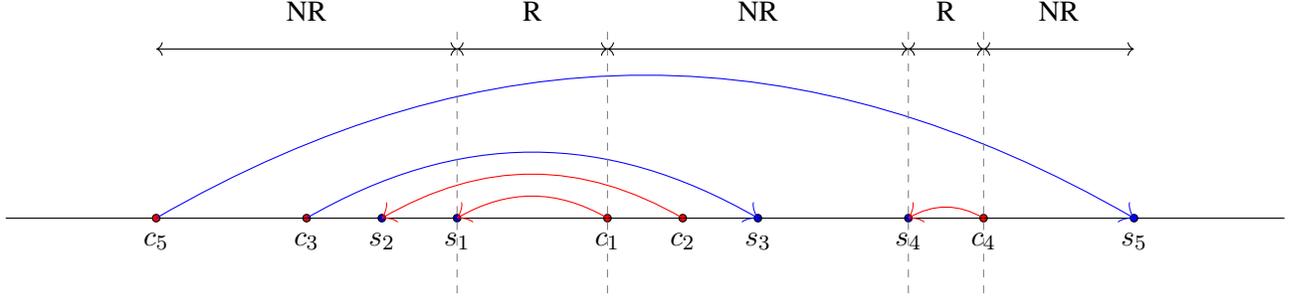
\begin{figure}
    \centering

\begin{tikzpicture}
\draw (0,0) node[label=below:{$c_3$},fill,circle, draw, fill=red,inner sep=0,minimum size=0.1cm] {};
\draw (5,0) node[label=below:{$c_2$},fill,circle, draw, fill=red,inner sep=0,minimum size=0.1cm] {};
\draw (4,0) node[label=below:{$c_1$},fill,circle, draw, fill=red,inner sep=0,minimum size=0.1cm] {};
\draw (9,0) node[label=below:{$c_4$},fill,circle, draw, fill=red,inner sep=0,minimum size=0.1cm] {};

\draw (2,0) node[label=below:{$s_1$},fill,circle, draw, fill=blue,inner sep=0,minimum size=0.1cm] {};
\draw (1,0) node[label=below:{$s_2$},fill,circle, draw, fill=blue,inner sep=0,minimum size=0.1cm] {};
\draw (6,0) node[label=below:{$s_3$},fill,circle, draw, fill=blue,inner sep=0,minimum size=0.1cm] {};
\draw (8,0) node[label=below:{$s_4$},fill,circle, draw, fill=blue,inner sep=0,minimum size=0.1cm] {};

\draw (-4,0) -- (13,0);

\draw [-{Classical TikZ Rightarrow[scale=2.5]},bend right,color=red](4,0) to (2,0);
\draw [-{Classical TikZ Rightarrow[scale=2.5]},bend right,color=red](5,0) to (1,0);
\draw [-{Classical TikZ Rightarrow[scale=2.5]},bend left,color=blue](0,0) to (6,0);
\draw [-{Classical TikZ Rightarrow[scale=2.5]},bend right,color=red](9,0) to (8,0);

\draw (-2,0) node[label=below:{$c_5$},fill,circle, draw, fill=red,inner sep=0,minimum size=0.1cm] {};
\draw (11,0) node[label=below:{$s_5$},fill,circle, draw, fill=blue,inner sep=0,minimum size=0.1cm] {};

\draw [-{Classical TikZ Rightarrow[scale=2.5]},bend left,color=blue](-2,0) to (11,0);

\draw [gray,dashed] (2,-1) -- (2,2.5);
\draw [gray,dashed] (4,-1) -- (4,2.5);
\draw [gray,dashed] (8,-1) -- (8,2.5);
\draw [gray,dashed] (9,-1) -- (9,2.5);

\draw [<->] (-2,2.25) to (2,2.25);
\draw [<->] (2,2.25) to (4,2.25);
\draw [<->] (4,2.25) to (8,2.25);
\draw [<->] (8,2.25) to (9,2.25);
\draw [<->] (9,2.25) to (11,2.25);

  \node at (0,2.75) 
{NR};
  \node at (3,2.75) 
{R};
  \node at (6,2.75) 
{NR};
  \node at (8.5,2.75) 
{R};
  \node at (10,2.75) 
{NR};

\end{tikzpicture}
    \caption{Illustration of redundant and non-redundant arcs. Suppose that the forward arc $(c_5,s_5)$ is added. The segments $[\ell(s_1),\ell(c_1)]$ and $[\ell(s_4),\ell(c_4)]$ of this arc are said to be \emph{redundant}, while others are \emph{non-redundant}.}
    \label{fig:redundant}
\end{figure}

We prove that our algorithm ensures that, if for an interval $I$, $n_t^f(I) > n_t^b(I) $, $n_t^f(I)-n_t^b(I)=disc_t(I)$ forward arc intervals are non-redundant for that interval, while the rest $n_t^b(I)$ forward arc intervals are redundant. 

\begin{lemma}
	\label{lem:redundant}
	For every integer $t>0$, after $t$ clients have arrived, in the execution of \recursivecancel, in any interval $I$, the number of redundant forward arc intervals is equal to the minimum of forward and backward arcs crossing $I$. 
\end{lemma}

\begin{proof}
	We prove the following two claims inductively on the number of client-server pairs added:
	\begin{enumerate}
		\item In any interval, the number of redundant forward arc intervals is equal to the minimum of the number of forward arcs and the number of backward arcs crossing this interval. 
		\item In any interval $I$, if there exist two forward arcs $a_1 =(c_1,s_1)$ and $a_2=(c_2,s_2)$ crossing the interval $I$ such that the arc interval of $a_1$ with respect to $I$ is redundant, and that of $a_2$ is non-redundant, then $\ell(s_2) \geq \ell(s_1)$. 
%		if both redundant and non-redundant forward arc intervals are present, then the server of the non-redundant arc is to the right of the server of redundant arc. 
	\end{enumerate}
	Let $c,s$ be the client-server pair given by \permutation.
	Consider the two cases, adding a forward arc and a backward arc:
	\begin{enumerate}
		\item Suppose that $s$ is to the right of $c$ i.e. the case when we add the forward arc directly. 
		If in an interval between $c$ and $s$, there are fewer forward arcs than backward arcs before adding $c$ and $s$, we mark that interval as a redundant interval. Observe that this ensures that in those intervals, redundant forward arc count increases, and is equal to the minimum of the number of forward and the number of backward arcs crossing the interval. 
		In intervals where the number of forward arcs is at least the number of backward arcs before adding $c$ and $s$, the interval is non-redundant. In this case, the minimum of forward and backward arcs does not increase, and thus the claim continues to remain valid. 
		
		For the second claim: If in an interval, the new arc is marked redundant, then using claim $1$ on the instance before adding the forward arc, we can infer that all the forward arcs in that interval are redundant. Thus, claim $2$ is void in this case. 
		If in an interval, the new arc is marked non-redundant, we need to show that the new server is to the right of any server whose arc is marked redundant. This follows directly from the fact that an unmatched server cannot be present in the middle of an arc (\Cref{lem:no-middle-server}), and hence, $s$ is to the right of the server of any forward arc that intersects $c,s$. 
		
		\item Suppose that $s$ is to the left of $c$ i.e. the case when we recursively add backward arc(s). 
		In this case, in an interval $I \in [\ell(s),\ell(c)]$, either a backward arc is added if there is no forward arc crossing $I$, or if there is at least one forward arc crossing $I$, the number of forward arcs crossing $I$ reduces by one. 
		The intervals outside $[\ell(s),\ell(c)]$ are not affected. 
		From Lemma~\ref{lem:shrink}, we know that the forward arcs corresponding to a server only shorten. 
		Thus, the second claim trivially follows. 
		
		The algorithm deletes a forward arc from the interval $I \in [\ell(s), \ell(c)]$ if there exists at least one forward arc crossing $I$ before adding the new client $c$. 
		If the number of forward arcs crossing $I$ is at most the number of backward arcs crossing $I$, then all the forward arc intervals in $I$ are labeled redundant, and we delete one such arc interval. 
		The property of claim $1$ still holds. 
		Similarly, the property holds if there are no backward arcs are crossing $I$ in which case, all the forward arc intervals are labeled non-redundant. 
		Thus, it remains to show that if there are both non-redundant and redundant arc intervals in $I$, our algorithm deletes the non-redundant arc interval. 
		We use claim $2$ here. If there are both redundant and non-redundant arc intervals in $I$, note that the server corresponding to the non-redundant arc is to the right of the server corresponding to the redundant arc. 
		
		Recall that there is at most one forward arc that is deleted from any interval. 
		If a redundant arc is deleted from an interval, when the arc is selected, it has the farthest server among all arcs that intersect the backward arc. This combined with the above fact implies that if a redundant arc is deleted from an interval, then there is no non-redundant arc in that interval. Thus, if there are both redundant and non-redundant forward arcs inside an interval, our algorithm deletes the non-redundant arc. 
	\end{enumerate}
\end{proof}

Thus, to bound the cost of the algorithm, it suffices to bound the cost of the redundant arc intervals, which we do below. 
\begin{lemma}
	\label{lem:suffix}
	If a forward arc $a=(c,s)$ is added during an iteration, then in any suffix of $a$, the length of non redundant intervals is at least the length of redundant intervals. 
\end{lemma}
\begin{proof}
	Let $A$ denote the set of clients and servers prior to adding $c$ and $s$. Let $x \in [\ell(c),\ell(s)]$. We are interested in the suffix $[x,\ell(s)]$ of the arc $(c,s)$. Introduce a virtual client $c'$ at $x$. Consider the optimal solution of $A \cup \{c',s\}$. 
	%Since we are following Permutation till before $c$ is added, the optimal cost of matching $A \cup \{c',s\}$ is at least the optimal cost of matching $A$.
	We claim that the optimal cost of matching $A \cup \{c',s\}$ is at least the optimal cost of matching $A$.
	Since \permutation is server optimal, the optimal cost of matching clients and servers of $A$ is at most that of matching clients and servers inside $A \cup \{s\}$ (leaving the extra server free). Since adding an extra client cannot decrease the optimal cost, the optimal cost of matching $A \cup \{c',s\}$ is at least that of $A \cup \{s\}$, which is at least that of $A$. 
	
	Recall that we can rewrite the optimal cost of a set of clients and servers $A$ as $\sum_{I} |I| |disc_A(I)|$. 
	When we add $\{c',s\}$ to $A$, the increase in the cost of optimal matching occurs precisely at the intervals where the number of clients to the left is greater than the number of servers (including $c',s$). And in other intervals in $[x,\ell(s)]$, the cost paid by the optimal matching decreases. 
	However, this exactly corresponds to the redundancy and non-redundancy of the forward arc $c,s$ within intervals inside $[x,\ell(s)]$. The intervals where the cost of optimal matching increases are the ones in which the arc is non-redundant, and the intervals where the cost of optimal matching decreases are the ones in which the arc is redundant. 
	As the optimal cost is non-decreasing, in any suffix $[x,\ell(s)]$, the sum of lengths of  
\end{proof}

%While we defer the proof to the appendix, we note that this step crucially uses the server-optimality of \permutation, and~\Cref{lem:shrink} of the leftward movement of forward arcs for \recursivecancel. For a quick intuition, we illustrate the proof with the toy example as used earlier. Consider an instance where $c_1$ arrives at location $1$ and \permutation matches it to a server at location $0$ with a backward arc. Then, suppose a client $c_2$ arrives at location $0$, and \permutation matches it with a server $s_2$ using a forward arc. Then, we can conclude that $s_2$ must be at location $\ell(s_2) \geq 2$, or else it would contradict the the server-optimality of \permutation. Using this property, we can infer that the interval $(0,1)$ which is redundant for arc $(c_2, s_2)$ is at most the length of $(1,\ell(s_2))$ the non-redundant interval for the arc.
%\Cref{lem:shrink} lets us use similar arguments for forward arcs which are obtained due to re-matches in~\cref{alg3:step10}. 

\begin{corollary}
	\label{cor:redundant-bound}
	For every $t>0$, after $t$ clients have arrived, the cost of redundant forward arc intervals is at most that of non-redundant forward arc intervals. 
\end{corollary}
\begin{proof}
	Summing up~\Cref{lem:suffix} over all the forward arcs gives us the required bound. 
\end{proof}

Finally, we can prove the bound on the competitive ratio of the \recursivecancel algorithm. 

\begin{theorem} \label{thm:cost-line}
	For every integer $t>0$, after $t$ clients have arrived, the cost of the matching of Algorithm~\ref{alg:recursivecancel} (\recursivecancel) is at most $3$ times the cost of the optimal offline matching $\mathcal{M}_t^*$. 
\end{theorem}
\begin{proof}
For the sake of analysis, for every interval $I$, let an arbitrary set of $\min(n^f_t(I), n^b_t(I))$ backward arcs be labeled redundant w.r.t this interval, and the rest to be non-redundant. Then, from~\Cref{lem:redundant}, the number of redundant backward arcs is equal to the number of redundant forward arcs in any interval.

For matching $\Mcal_t$, we denote the total cost of non-redundant forward arcs (resp. backward) as $\cost(\Mcal_t, NF)$ (resp. $\cost(\Mcal_t,NB)$). Similarly, we denote the total cost of redundant forward arcs (resp. backward) as $\cost(\Mcal_t, RF)$ (resp. $\cost(\Mcal_t,RB)$). 
Now, using this definition and from~\Cref{lem:redundant}, note that for any interval, we have that $\disc_t(I)$ is equal to the number of non-redundant arcs crossing $I$ (they will all either be forward or backward). Hence, we have that $\OPT = \cost(\Mcal^*_t) = \sum_{I} |I| \disc_t(I) = \cost(\Mcal_t, NF) + \cost(\Mcal_t, NB)$.  Moreover, the cost of $\Mcal_t$ maintained by \recursivecancel is at most
	$\cost(M_t, RF) + \cost(M_t, RB) + \cost(M_t, NF) + \cost(M_t , NB) = 2 \cost(M_t, RF) + \cost(M_t, NF) + \cost(M_t, NB) \leq 2 \cdot \cost(M_t, NF) + \cost(M_t, NF) + \cost(M_t, NB) \leq 3 \left( \cost(\Mcal_t, NF) + \cost(\Mcal_t, NB)\right)  \leq 3 \OPT$. The first equality is from the definition in the paragraph above and the first inequality is due to~\Cref{cor:redundant-bound}.  
\end{proof}

\subsection{Actual Algorithm \farthestserver}
\label{sec:linerecourse}

%\noindent\begin{minipage}{.5\textwidth}
\begin{algorithm}
\caption{Algorithm \farthestserver}\label{alg:farthestserver}
\begin{algorithmic}[1]
\For {each client $c_t$ arriving at time $t$ }
\LState let $s_t$ be the server \permutation matches $c_t$ to
\If {$(c_t,s_t)$ is a forward arc}
\LState {$\Mcal_{t} = \Mcal_{t-1} \cup (c_t,s_t)$}
\Else
\LState let $C_f$ denote all clients $c$ matched via forward arcs in $\Mcal_{t-1}$ with $\ell(c) \in [\ell(s_t), \ell(c_t)]$ 
\LState {let $S_f$ be the servers matched to $C_f$ in $\Mcal_{t-1}$}
\LState let $A_f \leftarrow C_f \cup S_f \cup \{c_t,s_t\} $
\LState let $\Mcal_{f}$ denote the matching between $C_f$ and $S_f$ in $\Mcal_{t-1}$
\LState $\Mcal_t = \Mcal_{t-1} \setminus \Mcal_f \cup$ \sweep$(A_f, \Mcal_f)$ \Comment{re-match all the clients and servers in $A_f$} \label{alg4:step10}
\EndIf
\EndFor
\end{algorithmic}
\end{algorithm}

%\end{minipage}%
%\begin{minipage}{.5\textwidth}
\begin{algorithm}
\caption{Algorithm \sweep ($A_f, \mold$) \Comment{re-match $A_f$ to cancel overlaps while minimally altering $\mold$}}
\begin{algorithmic}[1]
\LState set $\mnew = \emptyset$, list $L = \emptyset$
\For{ all $v \in A_f$ (in left to right order)}
	\If {$v$ is a server}
	\If {$L = \emptyset$}
	\LState $L = L \cup \{v\}$ 
	\ElsIf {$\mold(v) \in L$} \Comment{If $v$'s matching vertex in $\mold$ is present in $L$, match it}  
	\LState update $\mnew = \mnew \cup \{(\mold(v),v)\}$ and  $L = L \setminus \{v\} \setminus \{\mold(v)\}$
	\Else
	\LState let $c$ be client in $L$ with \emph{rightmost} (or) \emph{farthest} server 
	\LState update $\mnew = \mnew \cup \{(c, v)\}$ and $L = L \setminus \{c\} \setminus \{v\}$. %If $c$ is there in the list, it gets preference over other clients if the current match of the server is unavailable.
	\EndIf
    \Else \Comment{$v$ is a client}
	\If {$L \cap \Scal \neq \emptyset$} 
	\LState let $s = L \cap \Scal$ \Comment{intersection will be unique server by~\Cref{lem:unique-server}} 
	\LState update $\mnew = \mnew \cup \{(v, s)\}$ and $L = L \setminus \{s\} \setminus \{v\}$
	\Else
	\LState $L = L \cup \{v\}$
	\EndIf
	\EndIf
\EndFor
\end{algorithmic}
\end{algorithm}
%\end{minipage}

In~\Cref{app:bad-recourse}, we give an illustrative example where the \recursivecancel algorithm can have large recourse. In this section, we present our actual algorithm \farthestserver, which also \emph{maximally cancels} overlapping forward arcs whenever \permutation adds a backward arc but does it in a left-to-right sweep rather than a recursive loop, and within the sweep, the algorithm tries to \emph{preserve existing matched arcs} whenever possible, and chooses to ``cancel'' the forward arcs in a greedy manner.
That is, for every connected component $[l,r]$ of forward arcs intersecting with a backward arc that is being added, we pick the smallest number of forward arcs $(c_1,s_1), (c_2,s_2), \ldots, (c_k,s_k)$ such that the union of these arcs is $[l,r]$.

To be precise, we achieve this by the following greedy algorithm: 
First, we pick the forward arc $a$ starting at $l$. (Recall that we assume that all the clients are located at distinct points on the line). Next, we pick the forward arc that intersects $a$ that has the farthest server, set the new forward arc as $a$, and repeat till we cover the whole interval $[l,r]$. 
Once we obtain the set of forward arcs, we cancel them by replacing them with $(c_2, s_1), (c_3,s_2), \ldots, (c_k,s_{k-1})$, whereas $c_1$ and $s_k$ are now matched using backward arcs. 
\begin{figure}
    \centering

\begin{tikzpicture}
\draw (-1,0) node[label=below:{$c_1$},fill,circle, draw, fill=red,inner sep=0,minimum size=0.1cm] {};
\draw (1,0) node[label=below:{$c_2$},fill,circle, draw, fill=red,inner sep=0,minimum size=0.1cm] {};
\draw (2,0) node[label=below:{$c_3$},fill,circle, draw, fill=red,inner sep=0,minimum size=0.1cm] {};
\draw (5,0) node[label=below:{$c_4$},fill,circle, draw, fill=red,inner sep=0,minimum size=0.1cm] {};
\draw (6,0) node[label=below:{$c_5$},fill,circle, draw, fill=red,inner sep=0,minimum size=0.1cm] {};
\draw (12,0) node[label=below:{$c_6$},fill,circle, draw, fill=red,inner sep=0,minimum size=0.1cm] {};

\draw (4,0) node[label=below:{$s_1$},fill,circle, draw, fill=blue,inner sep=0,minimum size=0.1cm] {};
\draw (3,0) node[label=below:{$s_2$},fill,circle, draw, fill=blue,inner sep=0,minimum size=0.1cm] {};
\draw (7,0) node[label=below:{$s_3$},fill,circle, draw, fill=blue,inner sep=0,minimum size=0.1cm] {};
\draw (8,0) node[label=below:{$s_4$},fill,circle, draw, fill=blue,inner sep=0,minimum size=0.1cm] {};
\draw (10,0) node[label=below:{$s_5$},fill,circle, draw, fill=blue,inner sep=0,minimum size=0.1cm] {};
\draw (-3,0) node[label=below:{$s_6$},fill,circle, draw, fill=blue,inner sep=0,minimum size=0.1cm] {};

\draw (-4,0) -- (13,0);

\draw [-{Classical TikZ Rightarrow[scale=2.5]},bend right,color=red](12,0) to (-3,0);

\draw [-{Classical TikZ Rightarrow[scale=2.5]},bend left,color=violet](-1,0) to (4,0);
\draw [-{Classical TikZ Rightarrow[scale=2.5]},bend left,color=blue](1,0) to (3,0);
\draw [-{Classical TikZ Rightarrow[scale=2.5]},bend left,color=violet](2,0) to (7,0);
\draw [-{Classical TikZ Rightarrow[scale=2.5]},bend left,color=blue](5,0) to (8,0);
\draw [-{Classical TikZ Rightarrow[scale=2.5]},bend left,color=violet](6,0) to (10,0);

\draw [gray,dashed] (-1,-1) -- (-1,2.75);
\draw [gray,dashed] (10,-1) -- (10,2.75);

\draw [<->] (-1,2.5) to (10,2.5);

  \node at (4.5,2.9) 
{A connected component of forward arcs};

\end{tikzpicture}

\begin{tikzpicture}
\draw (-1,0) node[label=below:{$c_1$},fill,circle, draw, fill=red,inner sep=0,minimum size=0.1cm] {};
\draw (1,0) node[label=below:{$c_2$},fill,circle, draw, fill=red,inner sep=0,minimum size=0.1cm] {};
\draw (2,0) node[label=below:{$c_3$},fill,circle, draw, fill=red,inner sep=0,minimum size=0.1cm] {};
\draw (5,0) node[label=below:{$c_4$},fill,circle, draw, fill=red,inner sep=0,minimum size=0.1cm] {};
\draw (6,0) node[label=below:{$c_5$},fill,circle, draw, fill=red,inner sep=0,minimum size=0.1cm] {};
\draw (12,0) node[label=below:{$c_6$},fill,circle, draw, fill=red,inner sep=0,minimum size=0.1cm] {};

\draw (4,0) node[label=below:{$s_1$},fill,circle, draw, fill=blue,inner sep=0,minimum size=0.1cm] {};
\draw (3,0) node[label=below:{$s_2$},fill,circle, draw, fill=blue,inner sep=0,minimum size=0.1cm] {};
\draw (7,0) node[label=below:{$s_3$},fill,circle, draw, fill=blue,inner sep=0,minimum size=0.1cm] {};
\draw (8,0) node[label=below:{$s_4$},fill,circle, draw, fill=blue,inner sep=0,minimum size=0.1cm] {};
\draw (10,0) node[label=below:{$s_5$},fill,circle, draw, fill=blue,inner sep=0,minimum size=0.1cm] {};
\draw (-3,0) node[label=below:{$s_6$},fill,circle, draw, fill=blue,inner sep=0,minimum size=0.1cm] {};

\draw (-4,0) -- (13,0);

\draw [-{Classical TikZ Rightarrow[scale=2.5]},bend right,color=red](-1,0) to (-3,0);
\draw [-{Classical TikZ Rightarrow[scale=2.5]},bend right,color=red](12,0) to (10,0);

\draw [-{Classical TikZ Rightarrow[scale=2.5]},bend left,color=blue](1,0) to (3,0);
\draw [-{Classical TikZ Rightarrow[scale=2.5]},bend left,color=violet](2,0) to (4,0);
\draw [-{Classical TikZ Rightarrow[scale=2.5]},bend left,color=blue](5,0) to (8,0);
\draw [-{Classical TikZ Rightarrow[scale=2.5]},bend left,color=violet](6,0) to (7,0);
\end{tikzpicture}

    \caption{Illustration of farthest server algorithm: When the backward arc $(c_6,s_6)$ is added, in the connected component shown in the figure, the algorithm first picks the arc $(c_1,s_1)$ as $a$. The forward arc that intersects $a$ with farthest server is $(c_3,s_3)$, and similarly, the next forward arc selected is $(c_5,s_5)$. The resulting arcs after cancellation are shown below. }
    \label{fig:farthest-server}
\end{figure}
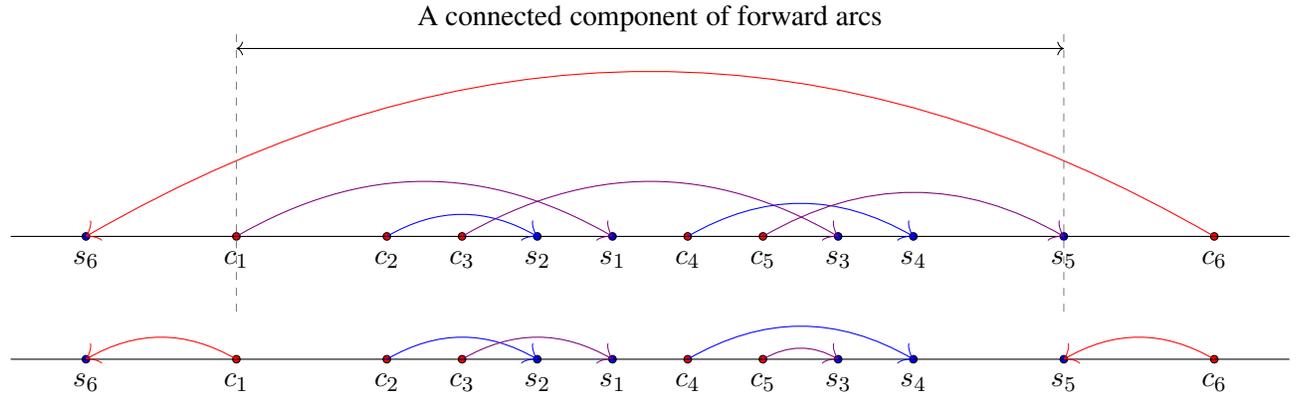
 We describe this in the \sweep procedure, which ensures that within a set $A_f$ of clients and servers where we need to re-match to cancel the backward arc, \emph{the new forward arcs of clients whose match have been changed are mutually disjoint}. 
This key property differentiates \farthestserver from \recursivecancel.
More formally, 
\begin{lemma}
	\label{lem:disjoint}
	When a new client $c_t$ arrives, the re-matched set of forward arcs \sweep$(A_f,\Mcal_f) \setminus \Mcal_f$ computed in ~\Cref{alg:farthestserver}, \cref{alg4:step10} are mutually disjoint. 
\end{lemma}

We state a couple of technical lemmas which help in proving Lemma~\ref{lem:disjoint}. 
%We defer the proofs to Appendix~\ref{app:line}.
\begin{lemma}
	\label{lem:shorten}
	If $c$ is a client that is currently matched to a server $s$ by a forward arc, and gets re-matched to server $s'$ during an iteration of the algorithm. Then $s'$ is to the left of $s$. 
\end{lemma}
\begin{proof}
	Suppose for the sake of contradiction that $s'$ is to the right of $s$. 
	Note that once a client gets matched via a backward arc, the algorithm does not rematch it. 
	Thus, $c$ is to the left of $s$, and thus to the left of $s'$. 
	As $c$ is in the list of clients when the algorithm considers $s'$, it is in the list when the algorithm considers $s$ as well. Thus, $s$ should not have been rematched. 
\end{proof}
\begin{lemma}
	\label{lem:unique-server}
	At any point during the execution of \sweep, the number of servers yet to be considered by the algorithm but whose client has been re-matched to a different server is at most one.  
\end{lemma}
\begin{proof}
	Fix the iteration as adding a backward arc $c,s$. We will prove the Lemma by inducting on the number of vertices (union of clients and servers) visited by the algorithm in the iteration. 
	Let $c'$ be the client matched with $s$ by the algorithm. 
	Note that $c'$ is the second vertex visited by the algorithm. 
	At this point, the server $s'$ that is originally matched to $c'$ is the only server satisfying the above property. 
	This covers the base case of the induction. 
	
	Suppose that the property is true till we add a new vertex $p$. 
	If $p$ is a client, adding $p$ to the list or matching it with a server to the left of it does not rematch any clients of servers yet unconsidered by the algorithm. 
	If $p$ is a server, if it is the unique server whose client has already been rematched, we now get a new server added to $list$. 
	If it has not lost its client yet, the set of servers whose clients are lost does not change (by design of the algorithm). 
\end{proof}

\begin{proofof}{\Cref{lem:disjoint}} Let $s$ be an arbitrary server that was re-matched during an iteration of the algorithm that is not the left-most server that is re-matched.
	Let $s_1$ be the server nearest to $s$ on the left side that is re-matched. 
	Note that such a server exists from the definition of $s$. 
	Let $c$ be the client matched to $s$ before re-matching. 
	We prove that the new matches of $s$ and $s_1$ are disjoint. 
	Applying this argument to all the servers other than the newly added server, we have proved the lemma. 
	
	We consider two cases separately. 
	\begin{OneLiners}
		\item $s_1$ now has a forward arc. 
		First, we claim that $c$ is the client that $s_1$ is re-matched to, after the end of the iteration. 
		Suppose for contradiction that $s_1$ is re-matched to $c' \neq c$. 
		Note that as $s$ is re-matched, $c$ should be re-matched too. As $s_1$ is the closest server on left to $s$ that is re-matched, $c$ should be re-matched to a server to the left of $s_1$. In other words, just after when we consider $s_1$ in the course of the algorithm, $s$ is a server whose client is already lost. However, note that the server originally matched to $c'$ has also now lost its client. Thus, we arrive at a contradiction to Lemma~\ref{lem:unique-server}.
		
		Let $c_1$ be the client that $s$ is re-matched to. We claim that $c_1$ is to the right of $s_1$ which proves that the new arcs adjacent to $s$ and $s_1$ are disjoint. 
		Suppose that $c_1$ is to the left of $s_1$. 
		From Lemma~\ref{lem:shorten}, we know that $c_1$'s original match is to the right of $s$. 
		Thus, when the algorithm is considering $s_1$, it would match $s_1$ with $c_1$ rather than $c$, since $c_1$ has the \emph{farthest server}.
		\item $s_1$ now has a backward arc. This case follows from the fact that the backward arcs that we added don't intersect with any other arcs. 
	\end{OneLiners}
	This completes the proof.
\end{proofof}

%Lemma~\ref{lem:disjoint} implies another simple but useful property similar to~\Cref{lem:shorten}: 
%\begin{lemma}
%	\label{lem:server-closer}
%	If a server $s$ has changed its match from $c_1$ to $c_2$, both using forward arcs, then $c_2$ lies to the right of $c_1$ i.e. $\ell(c_2) \geq \ell(c_1)$.
%\end{lemma}
%\begin{proof}
%	Suppose the server $s$ has changed its match from $c_1$ to $c_2$, and $c_2$ is to the left of $c_1$. 
%	As $c_1$ lies in between $c_2$ and $s$, the new arc of $c_1$ intersects $c_2$ and $s$ which directly contradicts Lemma~\ref{lem:disjoint}.
%\end{proof}
%We remark that~\Cref{lem:server-closer} is true for \recursivecancel as well, proved in~\Cref{lem:shrink}.

We now use the above lemmas to bound the recourse of the \farthestserver algorithm.

\begin{theorem}
	\label{thm:recourse}
	After the arrival of $k$ clients, the total recourse of \farthestserver algorithm is at most $\Ocal(k\log(k))$.
\end{theorem}

\begin{proof}
	Note that once a client is matched by a backward arc, it is not going to get re-matched later. 
	Thus, we are only interested in bounding the number of re-matches that match forward arcs to forward arcs. 
	\smallskip
	
	For a vertex (either client or server) $z$, let us define a ``length'' ${\rm len}(z)$ parameter which is equal to the number of vertices (all vertices which are part of the eventual matching after all $k$ clients have arrived) lying strictly inside the arc of $z$. 
	Therefore, before the start of the algorithm, the $len$ value of every vertex is at most $2k$. We now define the level of vertex $z$ as $\floor{\log {\rm len}(z)}$ so that the total initial level of all the vertices is at most $2k (1+\log k)$. Suppose that a client $c$ is currently matched to $s$ by using a forward arc, and in an iteration gets re-matched to $s'$ and $s$ gets re-matched to $c'$ both again using forward arcs. At least one of ${\rm len}(c)$ or ${\rm len}(s)$ should have decreased by at least a factor of $2$ since these are now disjoint arcs from~\Cref{lem:disjoint}. And therefore the total level of $c$ and $s$ at least decreases by 1. In other words, on every re-match, the total level decreases by at least 1, which together with the bound on the total initial level gives the number of such re-matches to be at most $2k ( 1+\log k)$.
	If at least one of $c$ or $s$ gets re-matched by a backward arc, its match does not change from then on. 
	Thus, the number of these type of re-matches are at most $2k$.
	Thus, in total, the recourse of the algorithm is $\Ocal(k \log k)$.
\end{proof}

\section{Lower bounds}

In this section, we present our lower bound for \omm on general metrics. 
\begin{theorem}
	Suppose that there exists an algorithm for \omm such that for every client $c$, the number of servers $s$ such that $c$ is matched to $s$ at some point of execution of the algorithm is at most $C$, for an absolute constant $C$. Then, the competitive ratio of the algorithm is at least $\Omega(\log(n))$. 
\end{theorem}

\begin{proof}
	We first describe the hard instance for \omm that we use to prove the lower bound. The underlying metric space is the star metric i.e. there exists a node $v_0$ that is at the center of the star, and a set of nodes $v_1, v_2, \ldots, v_n$ such that $d(v_0, v_i) =1 $ for all $i \in [n]$, and $d(v_i, v_j) = 2$ for all $i,j \in [n], i \neq j$. 
	For every $i \in [n]$, there is a server $s_i$ at $v_i$. 
	For each time $t=0,1, \ldots, n-1$ a single client $c_t$ arrives at a point in the metric space. 
	First, at $t=0$, the client $c_0$ arrives at $v_0$. 
	The next clients arrive at the location of the server just used by the algorithm. 
	Suppose that the algorithm matches $c_0$ to $s_i$. 
	Then, $c_1$ arrives at $v_i$. 
	After $t$ clients have arrived and have been matched by the algorithm, consider the server matched to $c_0$- let it be $s_{i_1}$. Let $s_{i_2}$ be the server matched by the algorithm to the client at $v_{i_1}$, and so on till there is no client yet arrived at $v_{i_k}$. Then, in our instance, at time $t$, a new client arrives at $v_{i_k}$. 

	Note that all the clients arrive at different locations in the metric space. This implies that at any point of time $t$, the offline optimal algorithm cost is equal to $1$. We can simply match each client $c_i$ other than $c_0$ to the server $s_i$, and match $c_0$ to an arbitrary unused server. 
	
	Suppose that for each client $c$, the number of servers $s$ such that $(c,s)$ is part of the matching of the algorithm at some point, is at most $C$. 
	Then, we claim that there is a time $t$ such that the online algorithm has cost at least $\Omega(\log (n))$ at time $t$. 
	Let $M_t$, $t=0,1,\ldots,n-1$ denote the matching maintained by the algorithm after time $t$.  	
	We consider a new algorithm that maintains a set of matchings $M_t'$, $t=0,1,\ldots,n-1$ after time $t$. 
	For every time $t$, we obtain $M_t'$ from $M_t$ as follows: Let $M = M_t$. 
	While there exists a client $c$ located at $v_i$ matched in $M$ to a server $s$ at $v_j \neq v_i$, but the server $s_i$ is not used in $M$, we rematch $c$ to $s_i$ in $M$. 
	Note that this process terminates in at most $n$ steps. 
	When this process can no longer proceed, we output $M_t' = M$. 
	The cost of the matching $M_t'$ is at most the cost of $M_t$, as every iteration of the above procedure only decreases the cost of the matching. 
	For every client $c$, the number of servers $s$ such that $c$ is matched to $s$ in some $M_t'$, is at most $C+1$. 
	Finally, the new algorithm that maintains the matchings $M_t'$ has a key property that at any time $t$: the matching $M_t'$ can be described as a path: $(c_0, s_{i_1}), (c_1, s_{i_2}), \ldots, (c_t,s_{i_{t+1}})$ such that $c_j$ and $s_{i_j}$ are at the same location. 
	
	We now claim that there exist some time $t$ such that the size of $M_t'$ is at least $\Omega(\log(n))$, which proves the required lower bound. 
	We define a directed graph $G = (V,E)$. The vertex set of the graph $V$ is equal to $\{0,1,\ldots, n\}$. 
	There is an edge from $i$ to $j$ if for some time $t$, the client $c_i$ is matched to the server $s_j$ in $M_t'$. 
	The out-degree of every node is at most $C+1$ in $G$. 
	We also define the graphs $G_0, G_1, \ldots, G_{n-1}$ as follows: The vertex set of $G_k$ is the same as $G$ for every $k$. 
	There is an edge from $i$ to $j$ in $G_k$ if client $c_i$ is matched to $s_j$ in $M_k'$. 
	It follows from the definitions that for every $k \in \{0,1,\ldots, n-1\}$, $G_i$ is a subgraph of $G$. 
	
	Note that for each $k$, the graph $G_k$ is a path that starts at $0$ and ends at the index of the location of the client $c_k$. 
	Thus, all the graphs $G_0, G_1, \ldots, G_{n-1}$ are different path subgraphs of $G$ all of which start at vertex $0$ and end at a different vertex in $G$. 
	As the out-degree of every vertex is at most $C+1$ in $G$, the number of distinct paths of length at most $l$ in $G$ starting at $0$ is at most $(C+1)^l$.  
	Thus, there should exist at least one path whose length is $\frac{\log(n)}{\log(C+1)} = \Omega(\log(n))$. 
	As the length of the subgraph $G_i$ denotes the cost of the matching $M_i'$, we get the required lower bound on the competitive ratio. 
\end{proof}
\section{Dynamic Online Matching}
\label{sec:fully-online}

In this section, we look at the dynamic version of the problem: at the beginning of the instance, a set of servers $S_0$ is available for matching. At time $t$: one of four possible events can happen, {\it (i) client arrival:} a new client request $c_t$ arrives which must be matched;  {\it (ii) server departure:} an existing server departs, potentially requiring the client it was matched with to be re-matched; {\it (iii) server arrival}; and {\it (iv) client departure}. The algorithm is allowed to re-match clients, incurring a recourse equal to the number of re-matches.

In comparison to previous sections, we no longer have the nested set of clients and servers since there are departures. However, for sake of clarity in presentation, we abuse notation and continue to let $C_t$ denote the set of active clients at time $t$, and let $\Scal_t$ denote the servers present at time $t$. We assume that $|\Scal_t| \geq |C_t|$ for all $t$ for feasibility. The online matching is denoted by $\Mcal_t$, and the optimal offline matching of $C_t$ to $\Scal_t$ by $\Mcal^*_t$. % Unlike the previous case,  in the dynamic case any of the four types of events can necessitate changes to existing matching. 

Our algorithm for the fully dynamic setting gives us a randomized $(\Ocal(\log n), \Ocal(\log \Delta))$-algorithm for the fully dynamic problem where $n$ is the number of points of the metric space and $\Delta$ is its aspect ratio. The algorithm proceeds in two steps: we first embed the metric space into a more structured \emph{hierarchically well-separated tree (HST)} metric while incurring an $\Ocal(\log n)$
loss in the competitive ratio; then in the second step, we show how we can maintain a constant-competitive matching with a recourse of $\Ocal(\log \Delta)$ on HSTs. The ideas are in fact very similar to~\cite{bansal2007log}, but we can obtain improved competitive ratios because we can re-match clients whereas the algorithm in~\cite{bansal2007log} cannot.

\medskip \noindent {\bf Step 1: Embedding to HSTs.} We reduce the problem for general metrics to a special class of tree metrics called Hierarchically Well-Separated Trees (HST) by using a classical result of \cite{fakcharoenphol2004tight} (\Cref{thm:FRT_theorem} below). Informally, an $\alpha$-HST embedding for $\alpha \geq 1$ is a tree where all the leaves are at the same depth $D$ (root being depth 1), and correspond exactly to the points in the metric space. A function $d_T()$ defines the length for the tree edges. All the edges at a given depth have the same length, and the lengths increase geometrically going from leaves to the root: length at depth $\ell$ is at least $\alpha$ times the length at level $\ell+1$. The tree distance $d_T$ between two nodes is the sum of edge lengths of the unique simple path connecting them.
%The following theorem shows that any general metric can be embedded into a distribution of HST's.

\begin{theorem}[\cite{fakcharoenphol2004tight}] 
\label{thm:FRT_theorem}
Given any metric $(\Xcal, d)$ on $n$ points, there is a probability distribution on $\alpha$-HST's such that: {\it (i)} For each tree $T$ in the support of the distribution, the leaves of $T$ correspond to the nodes of $\Xcal$, {\it (ii)} the tree distance is at least the metric distance: $d_T(x,y) \geq d(x,y)$ for all $x,y \in \Xcal$, and {\it (iii)} The expected tree distance satisfies: $\expct{d_T(x,y)} \leq \Ocal(\alpha \log n) d(x,y)$ for all $x,y \in \Xcal$.
\end{theorem}

 Setting $\alpha=2$ for our purposes, for embedding a metric with aspect ratio $\Delta$, Theorem~\ref{thm:FRT_theorem} gives a distribution of trees with depth at most $\Ocal(\log \Delta)$ such that all distances are preserved within a factor of  $\Ocal(\log n)$. %\cite{fakcharoenphol2004tight} also prove that the distribution can be efficiently sampled from, and further that the depth $D$ of the $\alpha$-HST is bounded by $\Ocal(\log \Delta)$ with probability 1. 
 
\medskip \noindent {\bf Step 2: Near-Optimal Algorithms on HSTs.}
We then show an $\Ocal(1)$-competitive algorithm on HSTs.
\begin{theorem}
\label{thm:dynamic_HST}
There exists a $(\Ocal(1), D)$-competitive deterministic algorithm for \ommr on an HST of depth $D$.
\end{theorem}

At a high level, we maintain the following invariant: for any sub-tree $T'$, the number of clients in $T'$ which have connections to servers outside $T'$ is equal to the excess number of clients over servers within $T'$. Algorithmically, when a new client $c$ arrives, it finds the smallest sub-tree $T'$ for which, either there is a free server $s$ in $T'$, or there is a server $s$ currently matched to a client $c'$ which is outside $T'$. In both cases, we match $c$ with $s$. Additionally, in the latter case, we re-match $c'$ using the same procedure, as if it were a new client. Server arrivals or departures can be handled similarly. 
%We defer the full details to Appendix~\ref{sec:dynamic_HST}.

We begin with a formal definition of Hierarchically Well-Separated trees (we borrow the following background on HST's and tree embeddings from \cite{bansal2007log}). 

\begin{definition}[Hierarchically Well-Separated Trees] An  $\alpha$-Hierarchically Well-Separated Tree for a given parameter $\alpha > 1$ is a rooted tree $T=(V,E)$ along with a length function $d()$ on the edges which satisfies the following properties:
	\begin{enumerate}
		\item For each node $v$, all its children are at the same distance from $v$.
		\item For any node $v$, $d(p(v), v) \geq \alpha d(v, c(v))$ where $p(v)$ is parent of $v$ and $c(v)$ is any child of $v$. That is, the length of the edges increase geometrically going from leaves to the root.
		\item Each leaf has the same distance to its parent.
	\end{enumerate}
\end{definition}

In this section, we will find convenient to talk about the {\it level} of nodes in the tree embedding of depth $D$. Our convention is that the root is at {\it level} $D$, and all the leaves are at level $1$. For a leaf $x$ matched to a leaf $y$ (one a client, and the other corresponding to a server)  the {\it level} of the match is defined as the level of the lowest common ancestor of $x$ and $y$ in the tree.

We describe the algorithm in~\Cref{alg:nearest-match}.
Recall that the nodes of our metric space are leaves of a $2$-HST.
Let $T$ be this underlying $2$-HST, and the distance between nodes $u$ and $v$ of $T$ is denoted by $d(u,v)$.
For a node $v$ of the tree, let $T(v)$ denote the subtree rooted at $v$.

In a nutshell, to make sure that solution cost is optimal, the algorithm ensures that for all nodes $v$ of the tree, the number of clients in $T(v)$ that are matched to servers outside $T(v)$ is given by the discrepancy between number of clients and servers in $T(v)$. 
\begin{algorithm}
	\caption{Nearest Match algorithm for Dynamic matching in HSTs}
	\label{alg:nearest-match}
	\begin{algorithmic}[1]
		\For {each time $t$}
		\If {client $c_t$ arrives at leaf $x$}
		\LState $level \leftarrow 1$
		\LState $c\leftarrow c_t$ \Comment{The unmatched client}
		\While {$c$ is unmatched}
		\LState Let $T_c$ be the subtree containing $c$ rooted at level $level$
		\If{ there is a free server $s$ in $T_c$}
		\LState Match $c,s$
		\LState Set level of $c$ and $s$ to be $level$ \Comment{Level of $c,s$ is set to level of LCA($c,s$).}
		\Else
		\If {$T_c$ has a server $s$ currently matched to a client $c'$ with level  $level' > level$}
		\LState Match $c,s$
		\LState Set level of $c$ and $s$ to be $level$
		\LState $c\leftarrow c'$
		\LState $level \leftarrow level'$
		\Else 
		\LState $level \leftarrow level + 1$
		\EndIf
		\EndIf
		\EndWhile
		\EndIf
		\If{server $s_t$ arrives at leaf $x$}
		\LState $level \leftarrow 1$
		\LState $s\leftarrow s_t$ \Comment{The unmatched server}
		\While{$level < D$}
		\LState $T_s$ is the subtree containing $s$
		\If{there is client $c \in T_s$ currently matched to server $s'$ with level $level'>level$}
		\LState Match $c,s$
		\LState $level \leftarrow level'$
		\LState $s\leftarrow s'$
		\Else 
		\LState $level \leftarrow level +1$
		\EndIf
		\EndWhile
		\EndIf
		\If{client $c_t$ departs}
		\LState Insert its currently matched server as a new server and run the second subroutine
		\EndIf
		\If{server $s_t$ departs}
		\LState Insert its currently matched client as a new client and run the first subroutine
		
		\EndIf
		\EndFor
	\end{algorithmic}
\end{algorithm}

\begin{proofof}{\Cref{thm:dynamic_HST}}
	In the algorithm, let us call an iteration to be running the algorithm when a new request arrives or departs.
	In any iteration, the $level$ parameter in the algorithm starts with a value equal to $1$ and increases whenever we change a match of a client or server.
	As the value of $level$ is upper bounded by the depth of the tree $D$, the number of changes in any iteration is at most $D$.
	
	To bound the cost of the algorithm, we use a proxy for the cost of a matching. 
	Suppose that a client $c$ is matched with server $s$ and let $p$ be the lowest common ancestor of $c$ and $s$ in $T$.
	Recall that $T$ satisfies the property that all edges between a node and its children are equal, and also that the distances at least double when we go up the tree.
	Using this, we can deduce that $d(p,s)\leq 2d(p,c)$, and thus $d(c,s)\leq 3d(p,c)$.
	Thus, by losing a factor of $3$, we can assume that the cost of matching $c$ and $s$ is, in fact, $d(c,p)$.
	This new cost is equivalent to summing over all $v$ in $T$, $d(v,p(v))$ times the number of clients in $T(v)$ that are matched to servers outside $T(v)$.
	
	We will prove that the algorithm has the property that at any point of time, for every node $v$ in the tree, in the subtree $T(v)$ rooted at $v$, the number of clients in $T(v)$ that are matched outside $T(v)$ is equal to the difference $\ell$ between the number of clients and servers in $T(v)$
	(If there are more servers than clients in $T(v)$, $\ell$ is set to be zero).
	Thus, the edge between $v$ and $p(v)$ ($p(v)$ is the parent of $v$) is used exactly $\ell$ times in our algorithm.
	Note that in any matching between clients and servers arrived so far, this edge has to be used at least $\ell$ times. 
	As this holds for every node $v$ in the tree, for every edge, any matching has to pay at least the cost that our algorithm pays.
	Thus, overall the cost of our algorithm is at most the optimal cost of matching existing clients and servers at any time. 
	
	In order to prove that the algorithm has this property, we will use induction on time $t$. 
	The property is trivially true when there are no requests. 
	Suppose that a new client $c_t$ arrives.
	The discrepancy $\ell$ of clients and servers is affected only for the ancestors of $c_t$ i.e. nodes on the path from $c_t$ to the root. 
	The parameter $level$ in the algorithm corresponds to the level of the ancestor that is currently under consideration. 
	We start with $level=1$ and might go all the way to $level=D$, corresponding to the ancestors of $c_t$, starting with $c_t$ to the root of the tree. 
	Let $v$ be the ancestor of $c_t$ at level $level$ in the tree. 
	If there are at least as many clients as servers in $T(v)$ before adding $c_t$, $c_t$ is forced to be matched outside $T(v)$ and thus the number of clients in $T(v)$ matched to servers outside $T(v)$ is one more than earlier.
	Also, in this case, the discrepancy $\ell$ of clients and servers increases by one, proving that the property holds for $v$. 
	
	However, if there are fewer clients than servers in $T(v),$ either there is a free server or there is a server that is matched outside $T(v)$.
	In either case, we match $c_t$ inside $T(v)$, and thus ensuring that the property still holds for $v$.
	If we match $c_t$ to a free server, the discrepancy of $v$ is still zero, and no client in $T(v)$ is matched to a server outside $T(v)$. 
	For the ancestors of $v$, their discrepancy is unaffected as we are adding both a server and a client in their subtrees.
	If we match $c_t$ to a server $s$ that is currently matched to a client $c$, $c$ is now free.
	We start the same process with $c$. However, as $c$ is currently matched to $s$ through $v'$ that is the least common ancestor of $c$ and $s$, for all the nodes $u$ in the path between $v'$ and $c_t$, the number of clients is strictly more than the number of servers.
	For these nodes, the discrepancy is unaffected, and the number of clients crossing their subtree is also unaffected as $c$ is going to be matched outside the subtree as well. 
	Thus the property still holds for these nodes. 
	For the nodes that are ancestors of $v'$, we continue the same process, ensuring that the property holds for them as well.
	
	The proof for server $s_t$ arrival is similar to the previous case.
	As before, the parameter $level$ corresponds to the level of the ancestor of $s_t$ currently under consideration, and let $v$ be the node that is the ancestor of $s_t$ at level $level$ in the tree. 
	If there are at least as many servers as that of clients in $T(v)$ before adding $s_t$, $v$'s discrepancy is unaffected by arrival of $s_t$, and we increase $level$ by one and move to $p(v)$. 
	However, if there are fewer servers than clients in $T(v)$, from the inductive hypothesis, at least a client in $T(v)$ is matched outside $T(v)$. 
	We now match one such client $c$ to $s_t$, thus decreasing the number of clients matched outside $T(v)$ by one, which exactly corresponds to the fact that the discrepancy of $v$ also decreases by one. 
	This forces a server $s$ to be unmatched, and we now look at the ancestors of $s$. 
	As before, we can start with the ancestor of $s$, $u$ that is the lowest common ancestor of $c$ and $s$ since the ancestors of $s$ below it are unaffected by $s$ becoming free. 
	Thus, the inductive property is preserved for all vertices when a server arrives. 
	
	Note that the property that the number of clients matched outside $T(v)$ is equal to the discrepancy $\ell$ between clients and servers in $T(v)$ is preserved when we remove a client and the server that it is matched to. 
	Thus, the property still holds when we delete a client or a server -- we can view it as first deleting a client-server pair and then adding the remaining client or server. 
\end{proofof}

Combining the two steps, we get the required algorithm for general metrics:
\begin{theorem}
	\label{thm:dynamicHST}
	For any $n$ point metric with aspect ratio $\Delta$, there exists a randomized online algorithm which is $( \Ocal(\log n),  \Ocal(\log \Delta))$-competitive for min-cost matching against an oblivious adversary.
\end{theorem}
\begin{proof}
	The algorithm proceeds by first sampling a $2$-HST and then executes the Algorithm from Theorem~\ref{thm:dynamic_HST}. Since the depth of the tree is $\Ocal(\log \Delta)$ with probability 1, this implies the recourse bound in the theorem. 
	
	To bound the cost, let $T$ denote the tree sampled during the construction of the $2$-HST, and let $\Mcal_t$ denote the matching produced by the algorithm (a random quantity). Recall that we denote the optimal matching at time $t$ by $\Mcal^*_t$. Let the cost of a matching $\Mcal$ under the metric induced by tree $T$ be denoted by $\cost_T$.
	
	By constant-factor optimality of $\Mcal_t$ for the particular HST sampled gives:
	\begin{align*}
	\cost_T(\Mcal_t) & \leq  C \cdot \cost_T(\Mcal^*_t).
	\end{align*}
	for some absolute constant $C$. Combined with the second property, this gives:
	\begin{align*}
	\cost(\Mcal_t) & \leq  C \cdot \cost_T(\Mcal^*_t).
	\end{align*}
	Taking expectation over the randomization of the algorithm,
	\begin{align*}
	\expct{\cost(\Mcal_t)} & \leq C \cdot \expct{ \cost_T(\Mcal^*_t)}
	\end{align*}
	which with $\expct{\cost_T( \Mcal_T)} \leq \Ocal(\log n) \cdot \cost(\Mcal_T)$ implies $\expct{\cost(\Mcal_t)} \leq \Ocal(\log n) \cdot \cost(\Mcal_t)$.
\end{proof}

\section{Conclusion and Open Questions}

The current work represents the first attempt at exploring the trade-off between the amount of recourse employed in online matching and the competitive ratio of the cost of the matching. However, we lack good lower bounds on the competitive ratio achievable. For example, for general metrics, it is unclear if $\Ocal(1)$ recourse is not enough to obtain ${\rm polylog}(k)$ competitive ratio. And for the line metric, we in fact conjecture that $(\Ocal(1), \Ocal(1))$-competitive online matching should be possible. It is even possible that \farthestserver has this property, but our current recourse analysis seems lacking. Another interesting avenue to explore is to limit the {\em worst-case} recourse used per time step and not just the average recourse, which is of practical concern. Finally, extending the results to random order or unknown {\it i.i.d.} models, and more broadly speaking, studying models where the algorithm can \emph{first select} the placement of servers, would be interesting.

\medskip \noindent {\bf Acknowledgements. } The authors would like to thank Janardhan Kulkarni for several useful discussions through the course of this work.

\appendix 

\section{Proofs for Section~\ref{sec:general-metrics}}
\label{sec:general-metrics-proofs}

\begin{proofof}{\Cref{prop:batch-permutation-tight}}
For simplicity, we prove the proposition for the case $d=2$. 
An alternate view of Algorithm \batchperm is the following: Let the leaves of a complete balanced tree of degree $d$ denote the $k$ client arrivals. Whenever the arrival of a client completes a subtree (that is, it is the rightmost leaf in some subtree), the matching for the clients and their currently matched serves is re-solved optimally.

Our lower bound instance will be on the line metric and consist of two parts: a \textit{core} instance and a \textit{auxiliary} instance. The subsequent client arrival will be chosen as the next arrival from either the core or the auxiliary instance so as to obtain a large recourse cost as we describe soon. The servers for the core instance will at locations $\pm 1, \pm 2, \pm 3, \ldots, \pm k/2$, and the servers for the auxiliary instance will be $k$ servers at location $10 k$. The client arrival sequence in the core instance will be $\epsilon, -1-\epsilon, 1+\epsilon, -2-\epsilon, 2+\epsilon, -3-\epsilon, 3+\epsilon, \ldots$; the client arrival sequence for the auxiliary instance is $10k, 10k, 10k, \ldots$.

Note that the above instance have been set up so that on the arrival of a client from the core instance, the server added by \permutation to the matching is also from the core instance, and similarly for a client from the auxiliary instance a server from the auxiliary instance is added. Further, the same is done by \OPT so that it suffices to study the cost and recourse for the arrivals in the core instance.

To decide whether the next client arrival happens from the core or the auxiliary sequence, we first check whether the arrival completes any subtree. If it does, denote the largest subtree it completes by $T$, and by $T_1, \ldots, T_d$ the $d$ subtrees of the root of $T$ (so that the new arrival is the rightmost leaf of $T_d$). If the number of core arrivals so far in $T_d$ is even, then the new arrival is also chosen from the core sequence. Otherwise the new arrival is chosen from the auxiliary chosen. 

We first prove the recourse bound. The sequence in which \permutation adds servers when the clients arrive from the core sequence is $1, -1, 2 , -2, \ldots$. In particular, the new client and server are added on the opposite sides of a central matching that is built online. The construction of the client arrival sequence ensures that when \batchperm resolves the optimal matching for subtree $T = (T_1, T_2)$ the number of client arrivals in $T_2$ is odd, and hence batch resolving ends up rematching all clients in $T_1 \cup T_2$ (except at most one). (In the general $d$ case we have to assume $d$ is odd, in which case it is easy to show that all the subtrees $T_1, T_2, \ldots, T_{d}$ have odd number of core clients, and thus a $\frac{d-1}{d}$ fraction of clients are rematched in the subtree $T$.)

To study cost, consider the matching immediately after the arrival of the $i$th client, and let $i = \sum_{j=0}^{\ell} d^{j} k_j $ ($0 \leq k_j \leq d-1$) denote the base $d$ representation of $i$. In particular, consider the case $k_j = 1$ for $1 \leq j \leq \ell$. The matching consists of one batch of $d^{\ell}$ clients each, followed by 1 batch of $d^{\ell-1}$ clients and so forth. The cost of \OPT is at most $i$. However, the cost of \batchperm is  $\Omega(i \cdot \ell) = \Omega( i \log_d i)$.

\end{proofof}

\section{Appendix: Bad Example for Recourse of \recursivecancel} \label{app:bad-recourse}
We illustrate the fact that \recursivecancel algorithm can have bad recourse in the following example: 
\begin{center}
\begin{tikzpicture}
\draw (0,0) node[label=below:{$c_1$},fill,circle, draw, fill=red,inner sep=0,minimum size=0.1cm] {};
\draw (1,0) node[label=below:{$c_2$},fill,circle, draw, fill=red,inner sep=0,minimum size=0.1cm] {};
\draw (2,0) node[label=below:{$c_3$},fill,circle, draw, fill=red,inner sep=0,minimum size=0.1cm] {};
\draw (3,0) node[label=below:{$c_4$},fill,circle, draw, fill=red,inner sep=0,minimum size=0.1cm] {};

\draw (6,0) node[label=below:{$s_1$},fill,circle, draw, fill=blue,inner sep=0,minimum size=0.1cm] {};
\draw (7,0) node[label=below:{$s_2$},fill,circle, draw, fill=blue,inner sep=0,minimum size=0.1cm] {};
\draw (8,0) node[label=below:{$s_3$},fill,circle, draw, fill=blue,inner sep=0,minimum size=0.1cm] {};
\draw (9,0) node[label=below:{$s_4$},fill,circle, draw, fill=blue,inner sep=0,minimum size=0.1cm] {};

\draw (-4,0) -- (13,0);

\draw [-{Classical TikZ Rightarrow[scale=2.5]},bend left,color=blue](0,0) to (6,0);
\draw [-{Classical TikZ Rightarrow[scale=2.5]},bend left,color=blue](1,0) to (7,0);
\draw [-{Classical TikZ Rightarrow[scale=2.5]},bend left,color=blue](2,0) to (8,0);
\draw [-{Classical TikZ Rightarrow[scale=2.5]},bend left,color=blue](3,0) to (9,0);

\draw (11,0) node[label=below:{$c_5$},fill,circle, draw, fill=red,inner sep=0,minimum size=0.1cm] {};
\draw (-2,0) node[label=below:{$s_5$},fill,circle, draw, fill=blue,inner sep=0,minimum size=0.1cm] {};

\draw [-{Classical TikZ Rightarrow[scale=2.5]},bend right,color=red](11,0) to (-2,0);

\end{tikzpicture}
\smallskip

\begin{tikzpicture}
\draw (0,0) node[label=below:{$c_1$},fill,circle, draw, fill=red,inner sep=0,minimum size=0.1cm] {};
\draw (1,0) node[label=below:{$c_2$},fill,circle, draw, fill=red,inner sep=0,minimum size=0.1cm] {};
\draw (2,0) node[label=below:{$c_3$},fill,circle, draw, fill=red,inner sep=0,minimum size=0.1cm] {};
\draw (3,0) node[label=below:{$c_4$},fill,circle, draw, fill=red,inner sep=0,minimum size=0.1cm] {};

\draw (6,0) node[label=below:{$s_1$},fill,circle, draw, fill=blue,inner sep=0,minimum size=0.1cm] {};
\draw (7,0) node[label=below:{$s_2$},fill,circle, draw, fill=blue,inner sep=0,minimum size=0.1cm] {};
\draw (8,0) node[label=below:{$s_3$},fill,circle, draw, fill=blue,inner sep=0,minimum size=0.1cm] {};
\draw (9,0) node[label=below:{$s_4$},fill,circle, draw, fill=blue,inner sep=0,minimum size=0.1cm] {};

\draw (-4,0) -- (13,0);

\draw [-{Classical TikZ Rightarrow[scale=2.5]},bend left,color=blue](0,0) to (6,0);
\draw [-{Classical TikZ Rightarrow[scale=2.5]},bend left,color=blue](1,0) to (7,0);
\draw [-{Classical TikZ Rightarrow[scale=2.5]},bend left,color=blue](2,0) to (8,0);

\draw (11,0) node[label=below:{$c_5$},fill,circle, draw, fill=red,inner sep=0,minimum size=0.1cm] {};
\draw (-2,0) node[label=below:{$s_5$},fill,circle, draw, fill=blue,inner sep=0,minimum size=0.1cm] {};

\draw [-{Classical TikZ Rightarrow[scale=2.5]},bend right,color=red](11,0) to (9,0);
\draw [-{Classical TikZ Rightarrow[scale=2.5]},bend right,color=red](3,0) to (-2,0);

\node[draw] at (-2,1) {Recursive cancel};

\end{tikzpicture}
\begin{tikzpicture}
\draw (0,0) node[label=below:{$c_1$},fill,circle, draw, fill=red,inner sep=0,minimum size=0.1cm] {};
\draw (1,0) node[label=below:{$c_2$},fill,circle, draw, fill=red,inner sep=0,minimum size=0.1cm] {};
\draw (2,0) node[label=below:{$c_3$},fill,circle, draw, fill=red,inner sep=0,minimum size=0.1cm] {};
\draw (3,0) node[label=below:{$c_4$},fill,circle, draw, fill=red,inner sep=0,minimum size=0.1cm] {};

\draw (6,0) node[label=below:{$s_1$},fill,circle, draw, fill=blue,inner sep=0,minimum size=0.1cm] {};
\draw (7,0) node[label=below:{$s_2$},fill,circle, draw, fill=blue,inner sep=0,minimum size=0.1cm] {};
\draw (8,0) node[label=below:{$s_3$},fill,circle, draw, fill=blue,inner sep=0,minimum size=0.1cm] {};
\draw (9,0) node[label=below:{$s_4$},fill,circle, draw, fill=blue,inner sep=0,minimum size=0.1cm] {};

\draw (-4,0) -- (13,0);

\draw [-{Classical TikZ Rightarrow[scale=2.5]},bend left,color=blue](0,0) to (6,0);
\draw [-{Classical TikZ Rightarrow[scale=2.5]},bend left,color=blue](1,0) to (7,0);
\draw [-{Classical TikZ Rightarrow[scale=2.5]},bend left,color=blue](3,0) to (8,0);

\draw (11,0) node[label=below:{$c_5$},fill,circle, draw, fill=red,inner sep=0,minimum size=0.1cm] {};
\draw (-2,0) node[label=below:{$s_5$},fill,circle, draw, fill=blue,inner sep=0,minimum size=0.1cm] {};

\draw [-{Classical TikZ Rightarrow[scale=2.5]},bend right,color=red](11,0) to (9,0);
\draw [-{Classical TikZ Rightarrow[scale=2.5]},bend right,color=red](2,0) to (-2,0);

\end{tikzpicture}

\begin{tikzpicture}
\draw (0,0) node[label=below:{$c_1$},fill,circle, draw, fill=red,inner sep=0,minimum size=0.1cm] {};
\draw (1,0) node[label=below:{$c_2$},fill,circle, draw, fill=red,inner sep=0,minimum size=0.1cm] {};
\draw (2,0) node[label=below:{$c_3$},fill,circle, draw, fill=red,inner sep=0,minimum size=0.1cm] {};
\draw (3,0) node[label=below:{$c_4$},fill,circle, draw, fill=red,inner sep=0,minimum size=0.1cm] {};

\draw (6,0) node[label=below:{$s_1$},fill,circle, draw, fill=blue,inner sep=0,minimum size=0.1cm] {};
\draw (7,0) node[label=below:{$s_2$},fill,circle, draw, fill=blue,inner sep=0,minimum size=0.1cm] {};
\draw (8,0) node[label=below:{$s_3$},fill,circle, draw, fill=blue,inner sep=0,minimum size=0.1cm] {};
\draw (9,0) node[label=below:{$s_4$},fill,circle, draw, fill=blue,inner sep=0,minimum size=0.1cm] {};

\draw (-4,0) -- (13,0);

\draw [-{Classical TikZ Rightarrow[scale=2.5]},bend left,color=blue](0,0) to (6,0);
\draw [-{Classical TikZ Rightarrow[scale=2.5]},bend left,color=blue](2,0) to (7,0);
\draw [-{Classical TikZ Rightarrow[scale=2.5]},bend left,color=blue](3,0) to (8,0);

\draw (11,0) node[label=below:{$c_5$},fill,circle, draw, fill=red,inner sep=0,minimum size=0.1cm] {};
\draw (-2,0) node[label=below:{$s_5$},fill,circle, draw, fill=blue,inner sep=0,minimum size=0.1cm] {};

\draw [-{Classical TikZ Rightarrow[scale=2.5]},bend right,color=red](11,0) to (9,0);
\draw [-{Classical TikZ Rightarrow[scale=2.5]},bend right,color=red](1,0) to (-2,0);

\end{tikzpicture}

\begin{tikzpicture}
\draw (0,0) node[label=below:{$c_1$},fill,circle, draw, fill=red,inner sep=0,minimum size=0.1cm] {};
\draw (1,0) node[label=below:{$c_2$},fill,circle, draw, fill=red,inner sep=0,minimum size=0.1cm] {};
\draw (2,0) node[label=below:{$c_3$},fill,circle, draw, fill=red,inner sep=0,minimum size=0.1cm] {};
\draw (3,0) node[label=below:{$c_4$},fill,circle, draw, fill=red,inner sep=0,minimum size=0.1cm] {};

\draw (6,0) node[label=below:{$s_1$},fill,circle, draw, fill=blue,inner sep=0,minimum size=0.1cm] {};
\draw (7,0) node[label=below:{$s_2$},fill,circle, draw, fill=blue,inner sep=0,minimum size=0.1cm] {};
\draw (8,0) node[label=below:{$s_3$},fill,circle, draw, fill=blue,inner sep=0,minimum size=0.1cm] {};
\draw (9,0) node[label=below:{$s_4$},fill,circle, draw, fill=blue,inner sep=0,minimum size=0.1cm] {};

\draw (-4,0) -- (13,0);

\draw [-{Classical TikZ Rightarrow[scale=2.5]},bend left,color=blue](1,0) to (6,0);
\draw [-{Classical TikZ Rightarrow[scale=2.5]},bend left,color=blue](2,0) to (7,0);
\draw [-{Classical TikZ Rightarrow[scale=2.5]},bend left,color=blue](3,0) to (8,0);

\draw (11,0) node[label=below:{$c_5$},fill,circle, draw, fill=red,inner sep=0,minimum size=0.1cm] {};
\draw (-2,0) node[label=below:{$s_5$},fill,circle, draw, fill=blue,inner sep=0,minimum size=0.1cm] {};

\draw [-{Classical TikZ Rightarrow[scale=2.5]},bend right,color=red](11,0) to (9,0);
\draw [-{Classical TikZ Rightarrow[scale=2.5]},bend right,color=red](0,0) to (-2,0);

\end{tikzpicture}
\smallskip

\begin{tikzpicture}
\draw (0,0) node[label=below:{$c_1$},fill,circle, draw, fill=red,inner sep=0,minimum size=0.1cm] {};
\draw (1,0) node[label=below:{$c_2$},fill,circle, draw, fill=red,inner sep=0,minimum size=0.1cm] {};
\draw (2,0) node[label=below:{$c_3$},fill,circle, draw, fill=red,inner sep=0,minimum size=0.1cm] {};
\draw (3,0) node[label=below:{$c_4$},fill,circle, draw, fill=red,inner sep=0,minimum size=0.1cm] {};

\draw (6,0) node[label=below:{$s_1$},fill,circle, draw, fill=blue,inner sep=0,minimum size=0.1cm] {};
\draw (7,0) node[label=below:{$s_2$},fill,circle, draw, fill=blue,inner sep=0,minimum size=0.1cm] {};
\draw (8,0) node[label=below:{$s_3$},fill,circle, draw, fill=blue,inner sep=0,minimum size=0.1cm] {};
\draw (9,0) node[label=below:{$s_4$},fill,circle, draw, fill=blue,inner sep=0,minimum size=0.1cm] {};

\draw (-4,0) -- (13,0);

\draw [-{Classical TikZ Rightarrow[scale=2.5]},bend left,color=blue](3,0) to (6,0);
\draw [-{Classical TikZ Rightarrow[scale=2.5]},bend left,color=blue](1,0) to (7,0);
\draw [-{Classical TikZ Rightarrow[scale=2.5]},bend left,color=blue](2,0) to (8,0);

\draw (11,0) node[label=below:{$c_5$},fill,circle, draw, fill=red,inner sep=0,minimum size=0.1cm] {};
\draw (-2,0) node[label=below:{$s_5$},fill,circle, draw, fill=blue,inner sep=0,minimum size=0.1cm] {};

\draw [-{Classical TikZ Rightarrow[scale=2.5]},bend right,color=red](11,0) to (9,0);
\draw [-{Classical TikZ Rightarrow[scale=2.5]},bend right,color=red](0,0) to (-2,0);

\node[draw] at (-2,1) {Farthest server};

\end{tikzpicture}
\end{center}
In this instance, there are four clients $c_1,c_2,c_3$ and $c_4$, $\ell(c_1)<\ell(c_2)<\ell(c_3)<\ell(c_4)$ currently matched using forward arcs to $s_1, s_2,s_3$ and $s_4$ respectively such that $\ell(c_4)<\ell(s_1)<\ell(s_2)<\ell(s_3)<\ell(s_4)$.
A new client $c_5$ arrives to the right of $s_4$ and \permutation outputs $s_5$ to the left of $c_1$ as the server. 
Since the arc $(c_5,s_5)$ is backward, the algorithms try to fix the matching. 
The \recursivecancel i.e.~\Cref{alg:recursivecancel} changes the matching completely to obtain $(c_2,s_1),(c_3,s_2),(c_4,s_3)$ as the new forward arcs, where as~\Cref{alg:farthestserver} changes only $c_4$'s matching and keeps $c_2$ and $c_3$ intact. 
If there are $k$ such forward arcs, and if $k$ backward arcs arrive, ~\Cref{alg:recursivecancel} has a recourse of $\Omega(k^2)$ where as~\Cref{alg:farthestserver} has only $O(k)$ recourse.

\bibliography{references}
\bibliographystyle{alpha}

\end{document}